\documentclass{article} 
\usepackage{nips14submit_e,times}
\usepackage{hyperref}
\usepackage{url}
\usepackage{algorithm}
\usepackage{algorithmic}

\usepackage{amsmath,amssymb,latexsym}
\usepackage{bm,cite,color,epsf,epsfig,graphicx,multirow,paralist,times}

\newtheorem{theorem}{Theorem}

\newtheorem{lemma}[theorem]{Lemma}

\newtheorem{proposition}[theorem]{Proposition}
\newtheorem{remark}[theorem]{Remark}

\title{Streaming, Memory Limited Algorithms for Community Detection}

\author{
Se-Young Yun 
\\
MSR-Inria\\
23 Avenue d'Italie, Paris 75013 \\
\texttt{seyoung.yun@inria.fr} \\
\And
Marc Lelarge \thanks{Work performed as part of MSR-INRIA joint research centre. M.L. acknowledges the support of the French Agence Nationale de la Recherche (ANR) under reference ANR-11-JS02-005-01 (GAP project).}  \\
Inria \& ENS \\
23 Avenue d'Italie, Paris 75013 \\
\texttt{marc.lelarge@ens.fr} \\
\AND
Alexandre Proutiere \thanks{A. Proutiere's research is supported by the ERC FSA grant, and the SSF ICT-Psi project.} \\
KTH, EE School / ACL \\
Osquldasv. 10, Stockholm 100-44, Sweden \\
\texttt{alepro@kth.se} \\
}

\nipsfinalcopy 

\begin{document}

\maketitle

\begin{abstract}
In this paper, we consider sparse networks consisting of a finite number of
non-overlapping communities, i.e. disjoint clusters, so that there is higher density within clusters than across clusters. Both the intra- and inter-cluster edge densities vanish when the size of the graph grows large, making the cluster reconstruction problem nosier and hence difficult to solve. We are interested in scenarios where the network size
is very large, so that the adjacency matrix of the graph is hard to manipulate and store. The data stream model in which columns of the adjacency matrix are revealed sequentially constitutes a natural
framework in this setting. For this model, we develop two novel clustering algorithms that extract the clusters asymptotically accurately. The first algorithm is {\it offline}, as it needs to store and keep the assignments of nodes to clusters, and requires a memory that scales linearly with the network size. The second algorithm is {\it online}, as it may classify a node when the corresponding column is revealed and then discard this information. This algorithm requires a memory growing sub-linearly with the network size. To construct these efficient streaming memory-limited clustering
algorithms, we first address the problem of clustering with partial
information, where only a small proportion of the columns of the
adjacency matrix is observed and develop, for this setting, a new spectral algorithm which is of independent interest.
\end{abstract}

\section{Introduction}

Extracting clusters or communities in networks have numerous applications and constitutes a fundamental task in many disciplines, including social science, biology, and physics. Most methods for clustering networks assume that  pairwise ``interactions'' between nodes can be observed, and that from these observations, one can construct a graph which is then partitioned into clusters. The resulting graph partitioning problem can be typically solved using spectral methods \cite{boppana1987eigenvalues, mcsherry2001spectral, dasgupta2006spectral, coja2010,chaudhuri2012spectral}, compressed sensing and matrix completion ideas \cite{chen2012, chatterjee2012}, or other techniques \cite{Jerrum1998155}. 

A popular model and benchmark to assess the performance of clustering
algorithms is the Stochastic Block Model (SBM) \cite{holland1983},
also referred to as the planted partition model. In the SBM, it is
assumed that the graph to partition has been generated randomly, by
placing an edge between two nodes with probability $p$ if the nodes
belong to the same cluster, and with probability $q$ otherwise, with
$q<p$. The parameters $p$ and $q$ typically depends on the network
size $n$, and they are often assumed to tend to 0 as $n$ grows large,
making the graph sparse. This model has attracted a lot of attention
recently. We know for example that there is a phase transition
threshold for the value of $\frac{(p-q)^2}{p+q}$. If we are below the
threshold, no algorithm can perform better than the algorithm randomly
assigning nodes to clusters \cite{decelle2011,mossel2012stochastic},
and if we are above the threshold, it becomes indeed possible to beat
the naive random assignment algorithm \cite{massoulie2013}. A
necessary and sufficient condition on $p$ and $q$ for the existence of
clustering algorithms that are asymptotically accurate (meaning that
the proportion of misclassified nodes tends to 0 as $n$ grows large)
has also been identified \cite{yun2014}. We finally know that spectral
algorithms can reconstruct the clusters asymptotically accurately as
soon as this is at all possible, i.e., they are in a sense optimal.

We focus here on scenarios where the network size can be extremely
large (online social and biological networks can, already today,
easily exceed several hundreds of millions of nodes), so that the
adjacency matrix $A$ of the corresponding graph can become difficult
to manipulate and store. We revisit network clustering problems under
memory constraints. Memory limited algorithms are relevant in the
streaming data model, where observations (i.e. parts of the adjacency
matrix) are collected sequentially. We assume here that the columns of
the adjacency matrix $A$ are revealed one by one to the algorithm. An
arriving column may be stored, but the algorithm cannot request it
later on if it was not stored. The objective of this paper is to
determine how the memory constraints and the data streaming model
affect the fundamental performance limits of clustering algorithms,
and how the latter should be modified to accommodate these
restrictions. Again to address these questions, we use the stochastic
block model as a performance benchmark. Surprisingly, we establish
that when there exists an algorithm with unlimited memory that
asymptotically reconstruct the clusters accurately, then we can devise an asymptotically accurate algorithm that
requires a memory scaling linearly in the network size $n$, except if
the graph is extremely sparse. This claim
is proved for the SBM with parameters $p=a{f(n)\over n}$ and
$q=b{f(n)\over n}$, with constants $a>b$, under the assumption that
$\log n\ll f(n)$. For this model, unconstrained
algorithms can accurately recover the clusters as soon as
$f(n)=\omega(1)$ \cite{yun2014}, so that the gap between
memory-limited and unconstrained algorithms is rather narrow. We
further prove that the proposed algorithm reconstruct the clusters
accurately before collecting all the columns of the matrix $A$, i.e.,
it uses {\it less} than one pass on the data.
We also propose an online streaming algorithm with sublinear memory
requirement. This algorithm output the partition of the graph in an
online fashion after a group of columns arrives. Specifically, if
$f(n)=n^\alpha$ with $0<\alpha<1$, our algorithm requires as little
as $n^\beta$ memory with $\beta>\max\left(1-\alpha, \frac{2}{3}
\right)$. To the best of our knowledge, our algorithm is the first
sublinear streaming algorithm for community detection. Although streaming
algorithms for clustering data streams have been analyzed
\cite{gmmo00}, the focus in this theoretical computer science
literature is on worst case graphs and on approximation performance which
is quite different from ours.

To construct efficient streaming memory-limited clustering algorithms,
we first address the problem of clustering with {\it partial
  information}. More precisely, we assume that a proportion $\gamma$
(that may depend on $n$) of the columns of $A$ is available, and we
wish to classify the nodes corresponding to these columns, i.e., the
observed nodes. We show that a necessary and sufficient condition for
the existence of asymptotically accurate algorithms is
$\sqrt{\gamma}f(n)=\omega(1)$.
We also show that to classify the observed nodes efficiently, a
clustering algorithm must exploit the information provided by the
edges between observed and unobserved nodes. We propose such an
algorithm, which in turn, constitutes a critical building block in the
design of memory-limited clustering schemes.

To our knowledge, this paper is the first to address the problem of community detection in the streaming model, and with memory constraints. Note that PCA has been recently investigated in the streaming model and with limited memory \cite{caramanis2013}. Our model is different, and to obtain efficient clustering algorithms, we need to exploit its structure.

\section{Models and Problem Formulation}

We consider a network consisting of a set $V$ of $n$ nodes. $V$ admits
a hidden partition of $K$ non-overlapping subsets $V_1,\ldots,V_K$,
i.e., $V=\bigcup_{k=1}^KV_k$. The size of community or cluster $V_k$
is $\alpha_k n$ for some $\alpha_k>0$. Without loss of generality, let
$\alpha_1 \le \alpha_2 \le \dots \le \alpha_K$. We assume that when
the network size $n$ grows large, the number of communities $K$ and
their relative sizes are kept fixed. To recover the hidden partition,
we have access to a $n\times n$ symmetric random binary matrix $A$ whose entries are independent and satisfy: for all $v, w\in V$, $\mathbb{P}[A_{vw}=1]=p$ if $v$ and $w$ are
in the same cluster, and $\mathbb{P}[A_{vw}=1]=q$ otherwise, with
$q<p$. This corresponds to the celebrated Stochastic Block Model
(SBM). If $A_{vw}=1$, we say that nodes $v$ and $w$ are connected, or
that there is an edge between $v$ and $w$. $p$ and $q$ typically
depend on the network size $n$. To simplify the presentation, we
assume that there exists a function $f(n)$ 
, and two constants $a>b$ such that $p=a{f(n)\over n}$ and
$q=b{f(n)\over n}$. This assumption on the specific scaling of $p$ and
$q$ is not crucial, and most of the results derived in this paper hold
for more general $p$ and $q$ (as it can be seen in the proofs).
For an algorithm $\pi$, we
denote by $\varepsilon^\pi(n)$ the proportion of nodes
that are misclassified by this algorithm. We say that $\pi$ is
asymptotically accurate if
$\lim_{n\to\infty}\mathbb{E}[\varepsilon^\pi(n)]=0$.
Note that in our setting, if $f(n) =O(1)$, there is a non-vanishing fraction of
isolated nodes for which no algorithm will perform better than a
random guess. In particular, no algorithm can be asymptotically
accurate.
Hence, we assume that $f(n)=\omega(1)$, which constitutes a necessary condition
for the graph to be asymptotically connected, i.e., the largest
connected component to have size $n-o(n)$. 

In this paper, we address the problem of reconstructing the clusters
from specific observed entries of $A$, and under some constraints
related to the memory available to process the data and on the way
observations are revealed and stored. 
More precisely, we consider the two following problems. 

{\bf Problem 1. Clustering with partial information.} We first
investigate the problem of detecting communities under the assumption
that the matrix $A$ is partially observable. More precisely, we assume
that a proportion $\gamma$ (that typically depend on the network size
$n$) of the columns of $A$ are known. The $\gamma n$ observed columns
are selected uniformly at random among all columns of $A$. Given these
observations, we wish to determine the set of parameters $\gamma$ and
$f(n)$ such that there exists an asymptotically accurate clustering
algorithm.

{\bf Problem 2. Clustering in the streaming model and under memory constraints.}
We are interested here in scenarios where the matrix $A$ cannot be stored entirely, and restrict our attention to algorithms that require memory less than $M$ bits. Ideally, we would like to devise an asymptotically accurate clustering algorithm that requires a memory $M$ scaling linearly or sub-linearly with the network size $n$. In the streaming model, we assume that at each time $t=1,\ldots,n$, we observe a column $A_v$ of $A$ uniformly distributed over the set of columns that have not been observed before $t$. The column $A_v$ may be stored at time $t$, but we cannot request it later on if it has not been explicitly stored. The problem is to design a clustering algorithm $\pi$ such that in the streaming model, $\pi$ is asymptotically accurate, and requires less than $M$ bits of memory. We distinguish offline clustering algorithms that must store the mapping between all nodes and their clusters (here $M$ has to scale linearly with $n$), and online algorithms that may classify the nodes when the corresponding columns are observed, and then discard this information (here $M$ could scale sub-linearly with $n$).

\section{Clustering with Partial Information}

In this section, we solve Problem 1. In what follows, we assume that $\gamma n=\omega(1)$, which simply means that the number of observed columns of $A$ grows large when $n$ tends to $\infty$. However we are typically interested in scenarios where the proportion of observed columns $\gamma$ tends to 0 as the network size grows large. Let $(A_v,v\in V^{(g)})$ denote the observed columns of $A$. $V^{(g)}$
is referred to as the set of {\it green} nodes and we denote by
$n^{(g)}=\gamma n$ the number of green nodes. $V^{(r)}=V\setminus
V^{(g)}$ is referred to as the set of {\it red} nodes. Note that we
have no information about the connections among the red nodes. For any
$k=1,\ldots,K$, let $V_k^{(g)}=V^{(g)}\cap V_k$, and
$V_k^{(r)}=V^{(r)}\cap V_k$. We say that a clustering algorithm $\pi$
classifies the green nodes asymptotically accurately, if the
proportion of misclassified green nodes, denoted by
$\varepsilon^\pi(n^{(g)})$, tends to 0 as the network size $n$ grows
large.

\subsection{Necessary Conditions for Accurate Detection}

We first derive necessary conditions for the existence of
asymptotically accurate clustering algorithms. As it is usual in this
setting, the hardest model to estimate (from a statistical point of
view) corresponds to the case of two clusters of equal sizes (see
Remark \ref{rem} below). Hence,
we state our information theoretic lower bounds, Theorems \ref{th:1}
and \ref{th:2}, for the special case
where $K=2$, and $\alpha_1=\alpha_2$. Theorem \ref{th:1} states that if the proportion of observed columns $\gamma$ is such that $\sqrt{\gamma}f(n)$ tends to 0 as $n$ grows large, then no clustering algorithm can perform better than the naive
 algorithm that assigns nodes to clusters randomly.

\begin{theorem}\label{th:1} 
Assume that $\sqrt{\gamma}f(n)=o(1)$. Then under any clustering algorithm $\pi$, the expected proportion of misclassified green nodes tends to 1/2 as $n$ grows large, i.e., $\lim\limits_{n\to\infty}\mathbb{E}[\varepsilon^\pi(n^{(g)})]=1/2$.
\end{theorem}

Theorem \ref{th:2} (i) shows that this condition is tight in the sense that
as soon as there exists a clustering algorithm that classifies the
green nodes asymptotically accurately, then we need to have $\sqrt{\gamma}f(n) =\omega(1)$.
Although we do not observe the connections among red nodes, we might ask
to classify these nodes through their connection patterns with green
nodes. Theorem \ref{th:2} (ii) shows that this is possible only if
$\gamma f(n)$ tends to infinity as $n$ grows large.

\begin{theorem}\label{th:2}
(i) If there exists a clustering algorithm that classifies the green nodes asymptotically accurately, then we have: $\sqrt{\gamma}f(n)=\omega(1)$.\\
(ii) If there exists an asymptotically accurate clustering algorithm (i.e., classifying all nodes asymptotically accurately), then we have: ${\gamma}f(n)=\omega(1)$.
\end{theorem}

\begin{remark}\label{rem}
Theorems \ref{th:1} and \ref{th:2} might appear restrictive as they
only deal with the case of two clusters of equal sizes. This is not
the case as we will provide in the next section an algorithm achieving
the bounds of Theorem \ref{th:2} (i) and (ii) for the general case
(with a finite number $K$ of clusters of possibly different sizes). In
other words, Theorems \ref{th:1} and \ref{th:2} translates directly in
minimax lower bounds thanks to the results we obtain in Section \ref{sec:algo}. 
\end{remark}

Note that as soon as $\gamma f(n)=\omega(1)$ (i.e. the mean
degree in the observed graph tends to infinity), then standard
spectral method applied on the squared matrix $A^{(g)}=(A_{vw}, v,w\in
V^{(g)})$ will allow us to classify asymptotically accurately the
green nodes, i.e., taking into account only the graph induced by the
green vertices is sufficient. However if $\gamma f(n) =o(1)$ then no
algorithm based on the induced graph only will be able to classify the
green nodes. Theorem \ref{th:2} shows that in the range of parameters
$1/f(n)^2\ll \gamma \ll 1/f(n)$, it is impossible to cluster
asymptotically accurately the red nodes but the question of clustering
the green nodes is left open. 

\subsection{Algorithms}\label{sec:algo}

In this section, we deal with the general case and assume that the
number $K$ of clusters (of possibly different sizes) is known.
There are two questions of interest: clustering green and red
nodes. It seems intuitive that red nodes can be classified only if we are able to
first classify green nodes. Indeed as we will see below, once the
green nodes have been classified, an easy greedy rule is optimal for
the red nodes. 

{\bf Classifying green nodes.} Our algorithm to classify green nodes
rely on spectral methods. Note that as suggested above, in the regime
$1/f(n)^2\ll \gamma \ll 1/f(n)$, any efficient
algorithm needs to exploit the observed connections between green and
red nodes.
We construct such an algorithm below. We should stress that our
algorithm does not require to know or estimate $\gamma$ or $f(n)$.

When from the observations, a red node $w\in V^{(r)}$ is connected to at most a single green node, i.e., if $\sum_{v\in V^{(g)}}A_{vw}\le 1$, this red node is useless in the classification of green nodes. On the contrary, when a red node is connected to two green nodes, say $v_1$ and $v_2$ ($A_{v_1w}=1=A_{v_2w}$), we may infer that the green nodes $v_1$ and $v_2$ are likely to be in the same cluster. In this case, we say that there is an {\it indirect edge} between $v_1$ and $v_2$. 

To classify the green nodes, we will use the matrix
$A^{(g)}=(A_{vw})_{v,w\in V^{(g)}}$, as well as the graph of indirect
edges. However this graph is statistically different from the
graphs arising in the classical stochastic block model. 
Indeed, when a red node is
connected to three or more green nodes, then the presence of indirect
edges between these green nodes are not statistically independent. To
circumvent this difficulty, we only consider indirect edges created
through red nodes connected to exactly two green nodes. Let $V^{(i)} = \{
v : v \in V^{(r)} ~\mbox{and}~\sum_{w\in V^{(g)}} A_{wv} = 2 \}$. We denote by $A'$
the $(n^{(g)}\times n^{(g)})$ matrix reporting the number of such indirect
edges between pairs of green nodes: for all $v, w\in V^{(g)}$,
$A_{vw}'= \sum_{z\in V^{(i)}} A_{vz}A_{wz}$. 

Our algorithm to classify the green nodes consists in the following steps: \\
Step 1. Construct the indirect edge matrix $A'$ using red nodes connected to two green nodes only.\\
Step 2. Perform a spectral analysis of matrices $A^{(g)}$ and
  $A'$ as follows: first trim $A^{(g)}$ and $A'$ (to remove nodes with
  too many connections), then extract their $K$ largest eigenvalues and the corresponding eigenvectors.\\
Step 3. Select the matrix $A^{(g)}$ or $A'$ with the largest normalized $K$-th largest eigenvalue. \\
Step 4. Construct the $K$ clusters $V_1^{(g)},\ldots,V_K^{(g)}$ based on the eigenvectors of the matrix selected in the previous step.

The detailed pseudo-code of the algorithm is presented in Algorithm
\ref{alg:indirect}. Steps 2 and 4 of the algorithm are standard
techniques used in clustering for the SBM, see e.g. \cite{coja2010}. The
algorithms involved in these Steps are presented in the supplementary
material (see Algorithms 4, 5, 6). Note that to extract the $K$ largest eigenvalues and the corresponding eigenvectors of a matrix, we use the power method, which is memory-efficient (this becomes important when addressing Problem 2). Further observe that in Step 3, the algorithm exploits the information provided by the red nodes: it selects, between the direct edge matrix $A^{(g)}$ and the indirect edge matrix $A'$, the matrix whose spectral properties provide more accurate information about the $K$ clusters. This crucial step is enough for the algorithm to classify the green nodes asymptotically accurately whenever this is at all possible, as stated in the following theorem:

\begin{algorithm}[t!] \small
   \caption{Spectral method with indirect edges}
   \label{alg:indirect}
\begin{algorithmic}
\STATE {\bfseries Input:} $A\in \{0,1\}^{|V|\times |V^{(g)}|}$, $V$,
$V^{(g)}$, $K$
\STATE $V^{(r) }\leftarrow  V \setminus V^{(g)}$
\STATE $V^{(i)} \leftarrow \{v : v \in V^{(r)} ~\mbox{and}~\sum_{w\in V^{(g)}} A_{wv} = 2 \}$
\STATE $A^{(g)} \leftarrow (A_{vw})_{v,w\in
  V^{(g)}}$ and $A'  \leftarrow (A_{vw}'=\sum_{z\in V^{(i)}} A_{vz}A_{wz})_{v,w\in
  V^{(g)}}$ 
\STATE $\hat{p}^{(g)} \leftarrow \frac{\sum_{v,w \in V^{(g)}}
  A^{(g)}_{vw}}{|V^{(g)}|^2}$ and $\hat{p}' \leftarrow \frac{\sum_{v,w \in V^{(g)}} A'_{vw}}{|V^{(g)}|^2}$
\STATE $Q^{(g)},\sigma^{(g)}_K , \Gamma^{(g)} \leftarrow$ Approx($A^{(g)}$, $\hat{p}^{(g)}$, $V^{(g)}$, $K$ )
and $Q',\sigma'_K , \Gamma' \leftarrow$ Approx($A'$, $\hat{p}'$, $V^{(g)}$, $K$ )
\IF{ $\frac{\sigma^{(g)}_K}{\sqrt{|V^{(g)}| \hat{p}^{(g)}}} \cdot 1_{\{ |V^{(g)}|
    \hat{p}^{(g)} \ge 50\}} \ge
  \frac{\sigma'_K}{\sqrt{|V^{(g)}| \hat{p}'}} \cdot 1_{\{ |V^{(g)}| \hat{p}' \ge 50\}}$}
   \STATE $(S_k)_{1\le k \le K} \leftarrow$ Detection
   ($Q^{(g)}, \Gamma^{(g)}, K$) 
\STATE Randomly place nodes in $V^{(g)}\setminus \Gamma^{(g)}$ to partitions $(S_k)_{k=1,\ldots,K}$
\ELSE 
   \STATE $(S_k)_{1\le k \le K} \leftarrow$ Detection
   ($Q',\Gamma', K$) 
\STATE Randomly place nodes in $V^{(g)}\setminus \Gamma'$ to partitions $(S_k)_{k=1,\ldots,K}$
\ENDIF
   \STATE {\bfseries Output:} $(S_k)_{1\le k \le K}$,
\end{algorithmic}
\end{algorithm}

\begin{theorem}\label{th:4}
When $\sqrt{\gamma} f(n) = \omega(1)$,
Algorithm~\ref{alg:indirect} classifies the green nodes asymptotically accurately.
\label{thm:spectral-method-ind}
\end{theorem}

In view of Theorem \ref{th:2} (i), our algorithm is optimal. It might
be surprising to choose one of the matrix $A^{(g)}$ or $A'$ and throw
the information contained in the other one. But the following simple calculation
gives the main idea. To simplify, consider the case $\gamma f(n)=o(1)$
so that we know that the matrix $A^{(g)}$ alone is not sufficient to
find the clusters. In this case, it is easy to see that the matrix
$A'$ alone allows to classify as soon as
$\sqrt{\gamma}f(n)=\omega(1)$. Indeed, the probability of getting an
indirect edge between two green nodes is of the order $(a^2+b^2)f(n)^2/(2n)$
if the two nodes are in the same clusters and $abf(n)^2/n$ if they
are in different clusters. Moreover the graph of indirect
edges has the same statistics as a SBM with these probabilities of
connection. Hence standard results show that spectral methods will
work as soon as $\gamma f(n)^2$ tends to infinity, i.e. the mean
degree in the observed graph of indirect edges tends to infinity.
In the case where $\gamma f(n)$ is too large (indeed $\gg \ln(f(n))$),
then the graph of indirect edges becomes too sparse for $A'$ to be
useful. But in this regime, $A^{(g)}$ allows to classify the green nodes.
This argument gives some intuitions about the full proof of Theorem
\ref{th:4} which can be found in the Appendix.

An attractive feature of our Algorithm \ref{alg:indirect} is that
it does not require any parameter of the model as input except the
number of clusters $K$. In particular, our algorithm selects
automatically the best matrix among $A'$ and $A^{(g)}$ based on their
spectral properties.

{\bf Classifying red nodes.} From Theorem \ref{th:2} (ii), in order to
classify red nodes, we need to assume that $\gamma f(n) = \omega(1)$. Under this assumption, the green nodes are well classified under Algorithm~\ref{alg:indirect}. To classify the red nodes accurately, we show that it is enough to greedily assign these nodes to the clusters of green nodes identified using Algorithm 1. More precisely, a red node $v$ is assigned to the cluster that maximizes the number of observed edges between $v$ and the green nodes of this cluster. The pseudo-code of this procedure is presented in Algorithm~\ref{alg:redpartition}.

\begin{algorithm}[t!]\small
   \caption{Greedy selections}
   \label{alg:redpartition}
\begin{algorithmic}
\STATE {\bfseries Input:} $A\in \{0,1\}^{|V|\times |V^{(g)}|}$, $V$,
$V^{(g)}$, $( S^{(g)}_k)_{1\le k \le K}$.
\STATE $V^{(r)} \leftarrow V \setminus V^{(g)}$ and $S_k \leftarrow S^{(g)}_k ,$ for all $k$
   \FOR{$v \in V^{(r)}$ }
   \STATE Find $k^{\star} = \arg \max_{k} \{\sum_{w \in S^{(g)}_k} A_{vw} / |
  S^{(g)}_k | \} $ (tie broken uniformly at random)
   \STATE $S_{k^{\star}} \leftarrow S_{k^{\star}} \cup \{ v \}$
    \ENDFOR
   \STATE {\bfseries Output:} $(S_k)_{1\le k \le K}$.
\end{algorithmic}
\end{algorithm}

\begin{theorem} When $\gamma f(n) = \omega(1)$, combining Algorithms \ref{alg:indirect} and \ref{alg:redpartition} yields an asymptotically accurate clustering algorithm.
\label{thm:clust-algor-red}
\end{theorem}

Again in view of Theorem \ref{th:2} (ii), our algorithm is optimal. To summarize our results about Problem 1, i.e., clustering with partial information, we have shown that:\\
(a) If $\gamma \ll 1/f(n)^2$, no clustering algorithm can perform better than the naive algorithm that assigns nodes to clusters randomly (in the case of two clusters of equal sizes).\\
(b) If $1/f(n)^2 \ll\gamma \ll 1/f(n)$, Algorithm~\ref{alg:indirect} classifies the green nodes asymptotically accurately, but no algorithm can classify the red nodes asymptotically accurately.\\
(c) If $1/f(n)\ll\gamma $, the combination of Algorithm~\ref{alg:indirect} and Algorithm~\ref{alg:redpartition} classifies all nodes asymptotically accurately.

\section{Clustering in the Streaming Model under Memory Constraints}

In this section, we address Problem 2 where the clustering problem has
additional constraints. Namely, the memory available to the algorithm
is limited (memory constraints) and each column $A_v$ of $A$ is observed only once, hence if
it is not stored, this information is lost (streaming model).

In view of previous results, when the entire matrix $A$ is available
(i.e. $\gamma =1$) and when there is no memory constraint, we know
that a necessary and sufficient condition for the existence of
asymptotically accurate clustering algorithms is that
$f(n)=\omega(1)$. 
Here we first devise a clustering algorithm adapted to the streaming
model and using a memory scaling linearly with $n$ that is
asymptotically accurate as soon as $\log(n)\ll f(n)$. Algorithms 1 and 2 are the building
blocks of this algorithm, and its performance analysis leverages the
results of previous section. We also show that our algorithm does not
need to sequentially observe all columns of $A$ in order to accurately
reconstruct the clusters. 
In other words, the algorithm uses strictly less than one pass on the
data and is asymptotically accurate.

Clearly if the algorithm is asked (as above) to output the full partition of the
network, it will require a memory scaling linearly with $n$, the size
of the output. 
However, in the streaming model, we can remove
this requirement and the algorithm can output the full partition
sequentially similarly to an online algorithm (however our algorithm
is not required to take an irrevocable action after the arrival of
each column but will classify nodes after a group of columns arrives). In this
case, the memory requirement can be sublinear. We present an
algorithm with a memory requirement which depends on the density of
the graph. In the particular case where $f(n)=n^{\alpha}$ with
$0<\alpha<1$, our algorithm requires as little as $n^{\beta}$ bits of memory
with $\beta>\max\left(1-\alpha,\frac{2}{3}\right)$ to accurately
cluster the nodes. Note that when the graph is very sparse
($\alpha\approx 0$), then the community detection
is a hard statistical task and the algorithm needs to gather a lot of columns so that
the memory requirement is quite high ($\beta\approx 1$). 
As $\alpha$ increases, the graph becomes denser and the statistical
task easier. As a result, our algorithm needs to look at smaller
blocks of columns and the memory requirement decreases. However, for
$\alpha \geq 1/3$, although the statistical task is much easier, our
algorithm hits its memory constraint and in order to store blocks with
sufficiently many columns, it needs to subsample each column. 
As a result, the memory requirement of our algorithm does not decrease
for $\alpha\ge 1/3.$

The main idea of our algorithms is to successively treat {\it blocks}
of $B$ consecutive arriving columns. Each column of a block is stored
in the memory. After the last column of a block arrives, we apply
Algorithm 1 to classify the corresponding nodes accurately, and we
then merge the obtained clusters with the previously identified
clusters. In the online version, the algorithm can output the
partition of the block and in the offline version, it stores this
result. 
We finally remove the stored columns, and proceed with the next
block. For the offline algorithm, after a total of $T$ observed
columns, we apply Algorithm 2 to classify the remaining nodes so that
$T$ can be less than $n$. The pseudo-code of the offline algorithm is
presented in Algorithm \ref{alg:streaming}.
Next we discuss how to tune $B$ and $T$ so that the classification is
asymptotically accurate, and we compute the required memory to
implement the algorithm.

\begin{algorithm}[t!]\small
   \caption{Streaming offline}
   \label{alg:streaming}
\begin{algorithmic}
\STATE {\bfseries Input:} $\{A_1, \dots, A_T \}$, $p$, $V$, $K$
\STATE {\bfseries Initial:} $N \leftarrow$ $n\times K$ matrix filled
with zeros and  $B \leftarrow
 \frac{n h(n)}{\min\{np,n^{1/3}\}\log n} $
\STATE {\bfseries Subsampling:} $A_t \leftarrow$ Randomly erase
entries of $A_t$ with probability $\max\{0, 1-\frac{n^{1/3}}{np} \}$ 
\FOR{$\tau =1${\bfseries to} $\lfloor \frac{T}{B} \rfloor$}
\STATE $A^{(B)} \leftarrow$ $n \times B$ matrix where $i$-th column is $A_{i+(\tau -1)B}$
\STATE $(S^{(\tau)}_k) \leftarrow$ Algorithm~\ref{alg:indirect}
$(A^{(B)},V, \{ (\tau -1)B +1, \dots, \tau B\} ,K)$
\IF {$\tau =1$}
\STATE $\hat{V}_k \leftarrow S^{(1)}_k$ for all $k$ and $N_{v,k}
\leftarrow \sum_{w \in S^{(1)}_k} A_{wv} $ for all $v\in V$ and $k$
\ELSE
\STATE $\hat{V}_{s(k)} \leftarrow \hat{V}_{s(k)} \cup S^{(\tau)}_{k}$ for all
$k$ where $s(k) = \arg\max_{1\le i\le K} \frac{ \sum_{v \in \hat{V}_i}
  \sum_{w \in  S^{(\tau)}_{k}}A_{vw}}{ |\hat{V}_i||
  S^{(\tau)}_{k}|}$
\STATE $N_{v,s(k)}
\leftarrow N_{v,s(k)} + \sum_{w \in S^{(\tau)}_{k}}  A_{wv}
$ for all $v\in V$ and $k$
\ENDIF 
\ENDFOR
\STATE {\bfseries Greedy improvement :} $\bar{V}_{k}
\leftarrow \{ v : k = \arg\max_{1\le i \le K} \frac{N_{v,i}}{|\hat{V}_i|} \}$ for all $k$
\STATE {\bfseries Output:} $(\bar{V}_k)_{1\le k \le K}$,
\end{algorithmic}
\end{algorithm}

{\bf Block size.} We denote by $B$ the size of a block. Let $h(n)$ be such that the block size is $B={h(n)n\over f(n)\log(n)}$.
Let $\bar{f}(n) = \min\{ f(n), n^{1/3} \}$ which represents the order
of the number of positive entries of each column after the subsampling
process.  According to Theorem \ref{th:4} (with $\gamma = B/n$), to
accurately classify the nodes arrived in a block, we just need that
$\frac{B}{n} \bar{f}(n)^2=\omega(1)$, which is equivalent to
$h(n)=\omega({\log(n)\over \min\{ f(n), n^{1/3} \}})$. Now the merging
procedure that combines the clusters found analyzing the current block
with the previously identified clusters uses the number of connections
between the nodes corresponding to the columns of the current block to
the previous clusters. The number of these connections must grow large
as $n$ tends to $\infty$ to ensure the accuracy of the merging
procedure. Since the number of these connections scales as
$B^2{\bar{f}(n)\over n}$, we need that ${h(n)^2}=\omega(\min\{ f(n), n^{1/3} \}{\log(n)^2\over
  n})$. Note that this condition is satisfied as long as $h(n)=\omega({\log(n)\over \min\{ f(n), n^{1/3} \}})$.

{\bf Total number of columns for the offline algorithm.} To accurately classify the nodes whose
columns are not observed, we will show that we need the total number
of observed columns $T$ to satisfy $T=\omega({n\over \min\{ f(n), n^{1/3} \}})$ (which is
in agreement with Theorem \ref{thm:clust-algor-red}).

{\bf Required memory for the offline algorithm.} To store the columns
of a block, we need $\Theta(nh(n))$
bits. To store the previously identified clusters, we need at most
$\log_2(K) n$ bits, and we can store the number of connections between
the nodes corresponding to the columns of the current block to the
previous clusters using a memory linearly scaling with $n$. Finally,
to execute Algorithm 1, the power method used to perform the SVD (see
Algorithm 5) requires the same amount of bits than that
used to store a block of size $B$. In summary, the required memory is
$M=\Theta(nh(n)+n)$.
\begin{theorem} Assume that $h(n)=\omega({\log(n)\over \min\{ f(n), n^{1/3} \}})$  and $T=\omega({n\over
    \min\{ f(n), n^{1/3} \}})$. Then with $M = \Theta(n h(n) + n)$ bits, Algorithm
  \ref{alg:streaming}, with block size $B={h(n)n\over \min\{ f(n), n^{1/3} \}\log(n)}$
  and acquiring the $T$ first columns of $A$, outputs clusters
  $\hat{V}_1,\ldots,\hat{V}_K$ such that with high probability, there
  exists a permutation $\sigma$ of $\{1,\ldots,K\}$ such that: $
  \frac{1}{n} \left|\bigcup_{1\le k \le K} \hat{V}_k \setminus
    V_{\sigma(k)} \right| = O\left( \exp(-cT{\min\{ f(n), n^{1/3}
      \}\over n}) \right) $ with a constant $c>0$.
\label{thm:streaming-with-green}
\end{theorem}

Under the conditions of the above theorem,
Algorithm~\ref{alg:streaming} is asymptotically accurate. Now if
$f(n)=\omega(\log(n))$, we can choose $h(n)=1$. Then Algorithm~\ref{alg:streaming} classifies nodes accurately and uses a memory linearly scaling with $n$. Note that increasing the number of observed columns $T$ just reduces the proportion of misclassified nodes. For example, if $f(n)=\log(n)^2$, with high probability, the proportion of misclassified nodes decays faster than $1/n$ if we acquire only $T=n/\log(n)$ columns, whereas it decays faster than $\exp(-\log(n)^2)$ if all columns are observed. 


\begin{algorithm}[t!]\small
   \caption{Streaming online}
   \label{alg:streamingon}
\begin{algorithmic}
\STATE {\bfseries Input:} $\{A_1, \dots, A_n \}$, $p$, $V$, $K$ 
\STATE {\bfseries Initial:} $B \leftarrow
 \frac{n h(n)}{\min\{np,n^{1/3}\}\log n} $ and $\tau^{\star} = \lfloor \frac{T}{B} \rfloor$
\STATE {\bfseries Subsampling:} $A_t \leftarrow$ Randomly erase
entries of $A_t$ with probability $\max\{0,1- \frac{n^{1/3}}{np} \}$ 
\FOR{$\tau =1${\bfseries to} $\tau^{\star}$}
\STATE $A^{(B)} \leftarrow$ $n \times B$ matrix where $i$-th column is $A_{i+(\tau -1)B}$
\STATE $(S_k)_{1\le k \le K} \leftarrow$ Algorithm~\ref{alg:indirect}
$(A^{(B)},V, \{ (\tau -1)B +1, \dots, \tau B\},K)$
\IF {$\tau =1$}
\STATE $\hat{V}_k \leftarrow S_k$ for all $k$ 
\STATE {\bfseries Output at $B$:} $(S_k)_{1\le k \le K}$
\ELSE
\STATE 
\STATE $s(k) \leftarrow \arg\max_{1\le i\le K} \frac{ \sum_{v \in \hat{V}_i}
  \sum_{w \in  S_{k}}A_{vw}}{ |\hat{V}_i|| S_{k}|}$ for all $k$
\STATE {\bfseries Output at $\tau B$:} $(S_{s(k)})_{1\le k \le K}$
\ENDIF
\ENDFOR
\end{algorithmic}
\end{algorithm}

Our online algorithm is a slight variation of the offline
algorithm. Indeed, it deals with the first block exactly in the same
manner and keeps in memory the partition of this first block. It then handles the successive blocks as the first block and merges the partition of these blocks with those of the first block as done in the offline algorithm for the second block. Once this is done, the online algorithm just throw all the information away except
the partition of the first block. 
\begin{theorem}
 Assume that $h(n)=\omega({\log(n)\over \min\{ f(n), n^{1/3} \}})$, then Algorithm
  ~\ref{alg:streamingon} with block size $B= \frac{h(n)n}{\min\{f(n),n^{1/3}\}\log n}$
  is asymptotically accurate (i.e., after one pass, the fraction of
  misclassified nodes vanishes) and requires $\Theta (n h(n))$ bits of
  memory.
\end{theorem}

\section{Conclusion}
We introduced the problem of community detection with partial
information, where only an induced subgraph corresponding to a fraction of the
nodes is observed. In this setting, we gave a necessary condition for
accurate reconstruction and developed a new spectral algorithm which extracts the clusters whenever this is at all possible. Building on this result, we considered the streaming, memory limited problem of community detection and developed algorithms able to asymptotically reconstruct the clusters with a memory requirement
which is linear in the size of the network for the offline version of
the algorithm and which is sublinear for its online
version. To the best of our knowledge, these algorithms are the first community
detection algorithms in the data stream model. 
The memory requirement of these algorithms is non-increasing in the
density of the graph and determining the optimal memory requirement is an interesting open problem.

\clearpage
\newpage
\bibliographystyle{abbrv}
\bibliography{reference}

\begin{thebibliography}{10}

\bibitem{boppana1987eigenvalues}
R.~B. Boppana.
\newblock Eigenvalues and graph bisection: An average-case analysis.
\newblock In {\em Foundations of Computer Science, 1987., 28th Annual Symposium
  on}, pages 280--285. IEEE, 1987.

\bibitem{chatterjee2012}
S.~Chatterjee.
\newblock Matrix estimation by universal singular value thresholding.
\newblock {\em arXiv preprint arXiv:1212.1247}, 2012.

\bibitem{chaudhuri2012spectral}
K.~Chaudhuri, F.~C. Graham, and A.~Tsiatas.
\newblock Spectral clustering of graphs with general degrees in the extended
  planted partition model.
\newblock {\em Journal of Machine Learning Research-Proceedings Track},
  23:35--1, 2012.

\bibitem{chen2012}
Y.~Chen, S.~Sanghavi, and H.~Xu.
\newblock Clustering sparse graphs.
\newblock In {\em Advances in Neural Information Processing Systems 25}, pages
  2213--2221. 2012.

\bibitem{coja2010}
A.~Coja-Oghlan.
\newblock Graph partitioning via adaptive spectral techniques.
\newblock {\em Combinatorics, Probability \& Computing}, 19(2):227--284, 2010.

\bibitem{dasgupta2006spectral}
A.~Dasgupta, J.~Hopcroft, R.~Kannan, and P.~Mitra.
\newblock Spectral clustering by recursive partitioning.
\newblock In {\em Algorithms--ESA 2006}, pages 256--267. Springer, 2006.

\bibitem{decelle2011}
A.~Decelle, F.~Krzakala, C.~Moore, and L.~Zdeborov{\'a}.
\newblock Inference and phase transitions in the detection of modules in sparse
  networks.
\newblock {\em Phys. Rev. Lett.}, 107, Aug 2011.

\bibitem{feige2005spectral}
U.~Feige and E.~Ofek.
\newblock Spectral techniques applied to sparse random graphs.
\newblock {\em Random Structures \& Algorithms}, 27(2):251--275, 2005.

\bibitem{gmmo00}
S.~Guha, N.~Mishra, R.~Motwani, and L.~O'Callaghan.
\newblock Clustering data streams.
\newblock In {\em 41st {A}nnual {S}ymposium on {F}oundations of {C}omputer
  {S}cience ({R}edondo {B}each, {CA}, 2000)}, pages 359--366. IEEE Comput. Soc.
  Press, Los Alamitos, CA, 2000.

\bibitem{holland1983}
P.~Holland, K.~Laskey, and S.~Leinhardt.
\newblock Stochastic blockmodels: First steps.
\newblock {\em Social Networks}, 5(2):109 -- 137, 1983.

\bibitem{Jerrum1998155}
M.~Jerrum and G.~B. Sorkin.
\newblock The metropolis algorithm for graph bisection.
\newblock {\em Discrete Applied Mathematics}, 82(1–3):155 -- 175, 1998.

\bibitem{massoulie2013}
L.~Massouli{\'e}.
\newblock Community detection thresholds and the weak ramanujan property.
\newblock {\em CoRR}, abs/1311.3085, 2013.

\bibitem{mcsherry2001spectral}
F.~McSherry.
\newblock Spectral partitioning of random graphs.
\newblock In {\em Foundations of Computer Science, 2001. Proceedings. 42nd IEEE
  Symposium on}, pages 529--537. IEEE, 2001.

\bibitem{caramanis2013}
I.~Mitliagkas, C.~Caramanis, and P.~Jain.
\newblock Memory limited, streaming {PCA}.
\newblock In {\em NIPS}, 2013.

\bibitem{mossel2012stochastic}
E.~Mossel, J.~Neeman, and A.~Sly.
\newblock Stochastic block models and reconstruction.
\newblock {\em arXiv preprint arXiv:1202.1499}, 2012.

\bibitem{rudelson2008}
M.~Rudelson and R.~Vershynin.
\newblock The littlewood--offord problem and invertibility of random matrices.
\newblock {\em Advances in Mathematics}, 218(2):600--633, 2008.

\bibitem{yun2014}
S.~Yun and A.~Proutiere.
\newblock Community detection via random and adaptive sampling.
\newblock In {\em COLT}, 2014.

\end{thebibliography}

\newpage
\appendix
\section{Algorithms}

We present below three algorithms that constitute building blocks of the main algorithms presented in the paper.

\begin{algorithm}[h!]\small
   \caption{Approx ($A$, $\hat{p}$, $V$, $K$ )}
   \label{alg:trim}
\begin{algorithmic}
  \STATE {\bfseries Input:} $A$, $\hat{p}$, $V$, $K$
\STATE $\ell^{\star} \leftarrow \max\{ 1, \lfloor |V| \exp(-|V|\hat{p}) \rfloor \})$
   \FOR{$v \in V$}
\STATE $x_v \leftarrow \sum_{w \in V} A_{vw}$
   \ENDFOR
\STATE $x^{\star} \leftarrow$  $\ell^\star$-th largest $x_v$
\STATE $\Gamma \leftarrow \{ v | x_v \le x^{\star},~ v\in V \}$
\STATE $\bar{A} \leftarrow (A_{vw})_{v,w\in \Gamma}$
\STATE $(Q, \sigma_{K} ) \leftarrow$  Power Method $(\bar{A},
\Gamma, K)$ (Algorithm~\ref{alg:pm})
\STATE {\bfseries Output:} $( Q , \sigma_K , \Gamma)$
\end{algorithmic}
\end{algorithm}

\begin{algorithm}[h!]\small
   \caption{Power Method ($A$, $V$, $K$ )}
   \label{alg:pm}
\begin{algorithmic}
\STATE {\bfseries Input:} $A$, $V$, $K$
\STATE {\bfseries Initial:} $Q_0 \leftarrow$ Randomly choose $K$
orthonormal vectors and $\tau^{\star} =\lceil \log |V| \rceil$
\FOR{$\tau=1$ {\bfseries to} $\tau^{\star} $}
\STATE $A Q_{\tau-1} = Q_{\tau} R_{\tau}$
\ENDFOR
\STATE $\sigma_{K} \leftarrow$ $K$-th largest singular value of $R_{\tau^{\star}}$
\STATE {\bfseries Output:} $( Q_{\tau^{\star}} , \sigma_K )$
\end{algorithmic}
\end{algorithm}

\begin{algorithm}[h!]\small
   \caption{Detection ($Q,V, K$)}
   \label{alg:detect}
\begin{algorithmic}
\STATE {\bfseries Input:} $Q,V, K$ (let $Q_v$ denote the low of $Q$ corresponding to $v$)
\FOR{$i=1$ {\bfseries to} $\log |V|$ }  
\STATE $X_{i,v} \leftarrow \{ w \in V :
\| Q_w  -Q_v\|^2 \le  \frac{i }{|V| \log |V|} \} $ 
\STATE $T_{i,0}\leftarrow \emptyset$
\FOR{$k=1$ {\bfseries to} $K$ }
\STATE $v_k^{\star} \leftarrow \arg \max_{v} | X_{i,v}\setminus \bigcup_{l=1}^{k-1} T_{i,l} |$ 
\STATE $T_{i,k} \leftarrow X_{i,v_k^{\star}} \setminus \bigcup_{l=1}^{k-1} T_{i,l} $ and $\xi_{i,k} \leftarrow \sum_{v \in T_{i,k}}  Q_v/ |T_{i,k}| .$
\ENDFOR
\FOR{$v \in V \setminus ( \bigcup_{k=1}^K T_{i,k} )$}
\STATE $k^{\star} \leftarrow \arg \min_{k} \| Q_v -\xi_{i,k}\|$ 
\STATE $T_{i,k^{\star}} \leftarrow T_{i,k^{\star}} \cup \{v\}$
\ENDFOR
\STATE $r_i  \leftarrow \sum_{k=1}^K \sum_{v \in T_{i,k}} \| Q_v -\xi_{i,k}\|^2$
\ENDFOR
\STATE $i^{\star} \leftarrow \arg \min_{i} r_i.$
\STATE $S_k\leftarrow T_{i^\star,k}$ for all $k$
\STATE {\bfseries Output:} $(S_k)_{k=1,\ldots,K}$.
\end{algorithmic}
\end{algorithm}

\section{Proofs}

In the following, we denote by $\lambda_i (X)$ the $i$-th largest
singular value of matrix $X$. 

\subsection{Proof of Theorem~\ref{th:1} }

{\bf Preliminaries.} In what follows, we denote by $\sigma^{(g)}\in \{-1,1\}^{n^{(g)}}$ a vector that represents the repartition of nodes in the two communities, i.e., nodes $v$ and $w$ belong to the same community if and only if $\sigma_v^{(g)}=\sigma_w^{(g)}$. We also denote by $\hat\sigma^{(g)}\in \{-1,1\}^{n^{(g)}}$ the estimate of $\sigma^{(g)}$ that a clustering algorithm could return.

We further introduce the following notation. For any $k>0$ and any two vectors $x,y\in\{-1,1\}^k$, we denote by $d_H(x,y) = \sum_{i=1}^k 1(x_i\neq y_i)$ the Hamming distance between $x$ and $y$ and define
\begin{eqnarray*}
d(x,y) = \frac{1}{k} \min\{d_H(x,y), d_H(x,-y)\}.
\end{eqnarray*}

For an estimate $\hat{\sigma}^{(g)}$ of $\sigma^{(g)}$, the quantity
$d(\hat{\sigma}^{(g)},\sigma^{(g)})$ is exactly the fraction of
misclassified green nodes. Hence if estimate $\hat{\sigma}^{(g)}$ is
obtained from algorithm $\pi$, we have
$\epsilon^\pi(n^{(g)})=d(\hat{\sigma}^{(g)},\sigma^{(g)})$.
Note that $d(\hat{\sigma}^{(g)},\sigma^{(g)})\leq 1/2$.

We first state key lemmas for this proof. Their proofs are postponed to the end of this section.
\begin{lemma}
For any $\alpha<1/2$ and estimate $\hat{\sigma}^{(g)}$, we have as $n^{(g)}\to \infty$
\begin{eqnarray*}
\mathbb{P}(d(\hat{\sigma}^{(g)},\sigma^{(g)})>\alpha) \geq 1-\frac{n^{(g)}-H(\sigma^{(g)}|A)}{n^{(g)}(1-H(\alpha))}+o(1),
\end{eqnarray*}
where $H(\alpha) = -\alpha\log \alpha -(1-\alpha)\log(1-\alpha)$ and
$H(\sigma^{(g)}|A)$ is the conditional entropy of $\sigma^{(g)}$
knowing $A$.\label{lem:7}
\end{lemma}
\begin{lemma}\label{lem:MI}
As $n^{(g)}\to \infty$, we have:
\begin{eqnarray*}
H(A) -H(A|\sigma^{(g)}) \leq o(n^{(g)})+O(n^{(g)}\gamma f(n)^2).
\end{eqnarray*}
\end{lemma}

From the definition of conditional entropy, we have
\begin{eqnarray}
H(\sigma^{(g)}|A) = H(\sigma^{(g)})-H(A) +H(A|\sigma^{(g)}) = n^{(g)}(1-o(1)),\label{eq:siga}
\end{eqnarray}
since $H(\sigma^{(g)}) = \log {n^{(g)}\choose
  n^{(g)}/2} \geq n^{(g)}-\frac{1}{2}\log 2n^{(g)}$ and we have $H(A) -H(A|\sigma^{(g)}) = o(n^{(g)})$ from Lemma~\ref{lem:MI}.
As soon as $n^{(g)}\to \infty$, putting \eqref{eq:siga} into Lemma~\ref{lem:7}, we see
that for any $\alpha<1/2$ and any estimate $\hat{\sigma}^{(g)}$,
\begin{eqnarray*}
\mathbb{P}(d(\hat{\sigma}^{(g)},\sigma^{(g)})>\alpha) \to 1.
\end{eqnarray*}
If
$\hat{\sigma}^{(g)}$ is a random guess, i.e. for each $v\in V^{(g)}$,
  $\hat{\sigma}^{(g)}_v$ is equal to $1$ or $-1$ with probability
  $1/2$ independently of the rest, then for any $\alpha<1/2$, as soon as $n^{(g)}\to \infty$, we
  have by the weak law of large numbers,
$\mathbb{P}(d(\hat{\sigma}^{(g)},\sigma^{(g)})>\alpha) \to 1$. 
Since we have
\begin{eqnarray*}
\mathbb{E}[\epsilon^\pi(n^{(g)})]\geq \alpha \mathbb{P}(d(\hat{\sigma}^{(g)},\sigma^{(g)})>\alpha),
\end{eqnarray*}
and $\alpha$ can be chosen as close to $1/2$ as desired, the
result follows.

\subsection{Proof of Lemma~\ref{lem:7}}
We define the event
$E=\{d(\hat{\sigma}^{(g)},\sigma^{(g)})>\alpha\}$ and $P_e$ its probability.
We have
\begin{eqnarray*}
H(E,\sigma^{(g)}|\hat{\sigma}^{(g)}) &=&
H(\sigma^{(g)}|\hat{\sigma}^{(g)})+\underbrace{H(E|\sigma^{(g)},\hat{\sigma}^{(g)})}_{0}\\
&=& H(E|\hat{\sigma}^{(g)})+H(\sigma^{(g)}|E,\hat{\sigma}^{(g)})\\
&\leq& H(P_e) + P_e\log{n^{(g)} \choose n^{(g)}/2} +(1-P_e)(n^{(g)}H(\alpha)+\log n^{(g)}),
\end{eqnarray*}
where the last inequality follows from $H(E|\hat{\sigma}^{(g)})\leq
H(E) = H(P_e)$ and the fact that
\begin{eqnarray*}
|\{\sigma^{(g)}, d(\sigma^{(g)}, \hat{\sigma}^{(g)}) \leq
\alpha\}|=\sum_{i=0}^{\alpha n^{(g)}} {n^{(g)}\choose i}\leq (n^{(g)}\alpha+1){n^{(g)}\choose
\alpha n^{(g)}}\leq n^{(g)}2^{n^{(g)}H(\alpha)}.
\end{eqnarray*}
Using $H(P_e)\leq 1$ and that ${n^{(g)}\choose n^{(g)}/2} \leq 2^{n^{(g)}}$ for
sufficiently large $n^{(g)}$, we get
\begin{eqnarray*}
P_e \geq \frac{H(\sigma^{(g)}|\hat{\sigma}^{(g)})-1-n^{(g)} H(\alpha)-\log
  n^{(g)}}{n^{(g)}(1-H(\alpha))-\log n^{(g)}}.
\end{eqnarray*}
The claim follows from the data processing inequality which ensures
$H(\sigma^{(g)}|\hat{\sigma}^{(g)})\geq H(\sigma^{(g)}|A)$.

\subsection{Proof of Lemma~\ref{lem:MI}}
Thanks to independence, we have
\begin{eqnarray*}
H(A) - H(A | \sigma^{(g)}) 
& = & H(A^{(g)}) - H(A^{(g)} | \sigma^{(g)})+H(A^{(r)}) - H(A^{(r)} |\sigma^{(g)})
\end{eqnarray*}
We first deal with the first term $H(A^{(g)}) - H(A^{(g)} |
\sigma^{(g)})$. By the concavity of $p\mapsto H(p)$, we have
$H(A^{(g)})\leq {n^{(g)}\choose 2} H\left(\frac{p+q}{2} \right)$ and $H(A^{(g)} | \sigma^{(g)}) = 2{n^{(g)}/2
  \choose 2} H(p) + (\frac{n^{(g)}}{2})^2 H(q)$.
Hence, we get
\begin{align*}
H(A^{(g)}) - H(A^{(g)} | \sigma^{(g)}) \le & {n^{(g)} \choose 2}
H(\frac{p+q}{2}) - 2{n^{(g)}/2
  \choose 2} H(p) - (\frac{n^{(g)}}{2})^2 H(q) \cr
=&\frac{(n^{(g)})^2}{4}\left( 2H(\frac{p+q}{2}) - H(p) - H(q) \right) + o(n^{(g)})
\cr
=&\frac{(n^{(g)})^2}{4} \left(p\log \frac{2p}{p+q} +q\log \frac{2q}{p+q}\right) \cr &+
\frac{(n^{(g)})^2}{4}\left((1-p)\log\frac{2-2p}{2-p-q}+(1-q)\log\frac{2-2q}{2-p-q}
\right)+ o(n^{(g)})\cr
\le&\frac{(n^{(g)})^2}{4} \left( \frac{(p-q)^2}{p+q} + \frac{(p-q)^2}{2-p-q}
\right) + o(n^{(g)})\\
=& o(n^{(g)}) + o(n^{(g)}\gamma f(n)).
\end{align*}

We denote by $A^{(r)}_v$ the row vector of
$A^{(r)}$ corresponding to $v\in V^{(r)}$.
For the second term,  by independence we have
\begin{eqnarray*}
H(A^{(r)}) - H(A^{(r)} |\sigma^{(g)}) = (n-n^{(g)})\big( H(A_v^{(r)}) -
H(A_v^{(r)} | \sigma^{(g)})\big).
\end{eqnarray*}
For a vector $x\in \{-1,1\}^{V^{(g)}}$ and $\sigma^{(g)}$, we define
$|x|^+ = \sum_{v\in V^{(g)}, \:\sigma^{(g)}_v=1}x_v$, $|x|^- =
\sum_{v\in V^{(g)}, \:\sigma^{(g)}_v=-1}x_v$ and $|x|=|x|^++|x|^-$.
For a given $\sigma^{(g)}$, we have
$$\mathbb{P}[ A_v^{(r)} = x | \sigma^{(g)}] = \zeta ( |x|^+ , |x|^- ) ,$$
where
$$\zeta (i,j) = \frac{((\frac{p}{1-p})^{i}(\frac{q}{1-q})^{j}+(\frac{p}{1-p})^{j}(\frac{q}{1-q})^{i})(1-p)^{\frac{m}{2}}(1-q)^{\frac{m}{2}}}{2} .$$
Since $\sigma^{(g)}$ is uniformly distributed,
\begin{align*}
\mathbb{P}[ A_v^{(g)} = x ] =& {n^{(g)} \choose n^{(g)}/2}^{-1}
\sum_{\sigma^{(g)}:\sum_{v\in V^{(g)}}\sigma_v^{(g)}=0} \mathbb{P}[
A_v^{(g)} = x | \sigma^{(g)}] \cr
=& {n^{(g)} \choose n^{(g)}/2}^{-1}  \sum_{i=0}^{|x|}{|x| \choose
  i}{n^{(g)}-|x| \choose n^{(g)}/2-i} \zeta(i,|x|-i)  = \eta(|x|),
\end{align*}
where
$$\eta(k) = \frac{\sum_{ i=0}^{k} {n^{(g)}/2 \choose i}{n^{(g)}/2 \choose k-i}
\zeta(i,k-i)}{\sum_{ i=0}^{k} {n^{(g)}/2 \choose i}{n^{(g)}/2 \choose k-i}} .$$
Since $\eta(0) = \zeta(0,0)$, $\eta(1) = \zeta(1,0)=\zeta(0,1)$, and
${n^{(g)} \choose k} \eta(k) = \sum_{i=0}^{ k} {n^{(g)}/2 \choose i}{n^{(g)}/2 \choose k-i}
\zeta(i,k-i),$
\begin{align*}
&H(A_v^{(r)}) - H( A_v^{(r)} | \sigma^{(g)} ) \cr 
&=-\sum_{k=0}^{ n^{(g)} }  {n^{(g)} \choose k}
\eta(k) \log \eta(k)   +\sum_{i=0}^{n^{(g)}/2}\sum_{j=0}^{ n^{(g)}/2 }  {n^{(g)}/2 \choose i}{n^{(g)}/2 \choose j}
\zeta(i,j) \log \zeta(i,j)  \cr
 &= \sum_{i=0}^{n^{(g)}/2}\sum_{j=0}^{ n^{(g)}/2 } {n^{(g)}/2 \choose i}{n^{(g)}/2 \choose j}
\zeta(i,j) \log \frac{\zeta(i,j)}{\eta(i+j)} \cr
& =\sum_{k=2}^{n^{(g)}} \sum_{i=0}^{k} 1_{i \le n^{(g)}/2} 1_{k-i \le n^{(g)}/2}  {n^{(g)}/2 \choose i}{n^{(g)}/2 \choose k-i}
\zeta(i,k-i) \log \frac{\zeta(i,k-i)}{\eta(k)} \cr
&\le \sum_{k=2}^{n^{(g)}} \sum_{i=0}^{k} 1_{i \le n^{(g)}/2} 1_{k-i \le n^{(g)}/2}  {n^{(g)}/2 \choose i}{n^{(g)}/2 \choose k-i}
\zeta(i,k-i) k \log (\frac{p}{q}) \cr 
&\le \sum_{2\le k \le n^{(g)}} (n^{(g)}p)^k k \log (\frac{a}{b}) \cr
& = O((n^{(g)})^2 p^2 ),\end{align*}
where the last equality stems from $n^{(g)}p = o(1)$.
Thus,
$$(n-m)\big( H(A_v^{(r)}) - H(A_v^{(r)} | \sigma^{(g)})\big) = O(n(n^{(g)})^2
p^2 )  = O(n^{(g)} \frac{n^{(g)}}{n} f(n)^2 ) = O(n^{(g)}\gamma f(n)^2),$$ and the lemma
follows since $f(n)\geq 1$ so that $f(n)^2\geq f(n)$.

\subsection{Proof of Theorem~\ref{th:2}}

In the remaining proofs, we use $m$ instead of $n^{(g)}$ to denote the
number of green nodes. We first consider case (i) with $\gamma=\Theta(1)$. In this case, a
necessary condition for the existence of an asymptotically accurate
clustering algorithm is that the fraction of green nodes outside the
largest connected component of the observed graph vanishes as $n\to
\infty$. This condition imposes that $f(n)\to \infty$.

We now consider case (i) with $m = o(n)$, i.e. $\gamma
= o(1)$. Denote
by $\Phi$ the true hidden partition $(V^{(g)}_1 , V^{(g)}_2)$ for
green nodes. Let $\mathbb{P}_{\Phi}$ be the probability measure
capturing the randomness in the observations assuming that the network
structure is described by $\Phi$. We also introduce a slightly
different structure $\Psi$. The latter is described by clusters
$V'^{(g)}_1 =V^{(g)}_1 \cup \{v_2 \}\setminus \{ v_1 \}$,
$V'^{(g)}_2=V^{(g)}_2 \cup \{v_1 \} \setminus \{ v_2 \}$ with
arbitrary selected $v_1\in V^{(g)}_1$ and $v_2 \in V^{(g)}_2$.

Let $\pi\in \Pi$ denote a clustering algorithm for green nodes with
output $(\hat{V}^{(g)}_1, \hat{V}^{(g)}_2)$, and let $\mathcal{E} =
\hat{V}^{(g)}_1 \bigtriangleup  V^{(g)}_1$ be the set
of misclassified nodes under $\pi$. Note that in general in our
proofs, we always assume without loss of generality that $|
\hat{V}^{(g)}_1 \bigtriangleup V^{(g)}_1 | \le |\hat{V}^{(g)}_1 \bigtriangleup V^{(g)}_{2}|$, so that the set of misclassified nodes is really
$\mathcal{E}$. Further define $\mathcal{B}=\{ v_1 \in \hat{V}^{(g)}_1\}$ as the set of events where node $v_1$ is
correctly classified. We have $\varepsilon (m) = |\mathcal{E}|/m$.

Let $x_{i,j}$ be equal to one if there is an edge between nodes $i$
and $j$ and zero otherwise.
With a slight abuse of notation, we define the boolean functions $p(\cdot)$ and
$q(\cdot)$ as follows: $p(1) = af(n)/n=p$, $q(1)=bf(n)/n=q$ and
$p(0)=1-p(1)$, $q(0)=1-q(1)$. We introduce $L$ (a quantity that resembles the log-likelihood ratio
between $\mathbb{P}_\Phi$ and $\mathbb{P}_{\Psi}$) as:

\begin{align*}
L =& \sum_{i \in V_1'^{(g)}}\log \frac{q(x_{i,v_1})p(x_{i,v_2})}{p(x_{i,v_1})q(x_{i,v_2})} + 
\sum_{i \in V_2'^{(g)}}\log
\frac{p(x_{i,v_1})q(x_{i,v_2})}{q(x_{i,v_1})p(x_{i,v_2})} \cr
&+ \sum_{v \in V^{(r)}} \log \frac{\prod_{i \in
    V_1'^{(g)}}p(x_{v,i})\prod_{i \in V_2'^{(g)}}q(x_{v,i}) + \prod_{i \in V_1'^{(g)}}q(x_{v,i})\prod_{i \in V_2'^{(g)}}p(x_{v,i})}{\prod_{i \in
    V_1^{(g)}}p(x_{v,i})\prod_{i \in V_2^{(g)}}q(x_{v,i}) + \prod_{i \in V_1^{(g)}}q(x_{v,i})\prod_{i \in V_2^{(g)}}p(x_{v,i})},
\end{align*}
In what follows, we establish a relationship between $\mathbb{E} [\varepsilon (m) ]$ and $L$. For any function $g(m)$, 
\begin{eqnarray} 
\mathbb{P}_{\Psi} \{ L \le  g(m) \}& =& \mathbb{P}_{\Psi} \{ L \le g(m) , \bar{\mathcal{B}} \} + \mathbb{P}_{\Psi} \{ L \le g(m) , \mathcal{B} \}. 
\end{eqnarray}
We have: 
\begin{eqnarray}\mathbb{P}_{\Psi} \{ L \le  g(m) , \bar{\mathcal{B}}\} &= &
  \int_{\{ L \le  g(m) , \bar{\mathcal{B}}\}} d\mathbb{P}_{\Psi} \cr
&= & \int_{\{ L \le  g(m) , \bar{\mathcal{B}}\}} \prod_{i \in
  V_1'}\frac{\nu(x_{i, v_1})\nu(x_{i,v_2})}{p(x_{i,v_1})q(x_{i,v_2})}
\prod_{i \in V_2'}\frac{\nu(x_{i,v_1})\nu(x_{i,v_2})}{q(x_{i, v_1})p(x_{i,v_2})} d\mathbb{P}_{\Phi} \cr
&\le & \exp (g(m))\mathbb{P}_{\Phi} \{ L \le  g(m) , \bar{\mathcal{B}}\}~ \le ~ \exp (g(m))\mathbb{P}_{\Phi} \{ \bar{\mathcal{B}}\}\cr
&\le &2\exp (g(m))\mathbb{E}_{\Phi} [ \varepsilon (m)], \label{eq:chofm-1}
\end{eqnarray}
where the last inequality comes from the fact that,
\begin{eqnarray*}\mathbb{P}_{\Phi} \{ \mathcal{B} \} &\ge& 1- \mathbb{P}_{\Phi} \{ v_1 \notin
\hat{V}^{(g)}_1 \} \ge 1-2 \mathbb{E}_{\Phi}[ \varepsilon(n) ]. \end{eqnarray*}
We also have: 
\begin{equation}\mathbb{P}_{\Psi} \{ L \le  g(m) , \mathcal{B} \} \le
  \mathbb{P}_{\Psi} \{  \mathcal{B} \} \le
2 \mathbb{E}_{\Psi}[ \varepsilon(m) ]. \label{eq:chofm-2}
\end{equation} 
By \eqref{eq:chofm-1} and \eqref{eq:chofm-2}
$$\mathbb{P}_{\Psi} \{ L \le  g(n) \} \le 2\mathbb{E}_{\Phi}[\varepsilon(n) ]\exp(g(n)) +
2 \mathbb{E}_{\Psi}[ \varepsilon(n) ]. $$
Since  $\mathbb{E}_{\Phi}[\varepsilon(n) ]
=\mathbb{E}_{\Psi}[\varepsilon(n) ] =\mathbb{E} [\varepsilon(n) ]$ and
$\mathbb{E} [\varepsilon(n) ] = o(1)$,
 choosing $g(m) = \log
\left( \frac{1 }{8 \mathbb{E} [ \varepsilon(m) ]} \right)$, we obtain:\begin{equation}
 \lim\inf_{m \rightarrow \infty}\mathbb{P}_{\Psi} \{ L \ge  \log \left( \frac{1}{8 \mathbb{E} [ \varepsilon(m) ]} \right) \} > \frac{1}{2}. 
\end{equation}
By Chebyshev's inequality, $\mathbb{P}_{\Psi} \{ L \ge
\mathbb{E}_{\Psi}[L] + 2\sigma_{\Psi}[L] \} \le
\frac{1}{4}$. Therefore, to be valid the above inequality,
\begin{equation}
\mathbb{E}_{\Psi}[L] + 2\sigma_{\Psi}[L] \ge \log \left(\frac{1}{8 \mathbb{E} [ \varepsilon(m) ]} \right),
\end{equation}
which implies that $\mathbb{E}_{\Psi}[L] + 2\sigma_{\Psi}[L] =
\omega(1)$ since $\mathbb{E} [ \varepsilon(m) ] = o(1)$.

We define $KL(p,q)=p\log(p/q)+(1-p)\log((1-p)/(1-q))$.
From the definition of $L$, we can easily bound $\mathbb{E}_{\Psi}[L]$
and $\sigma_{\Psi}[L]^2$ :
\begin{align}
\mathbb{E}_{\Psi}[L] \le& m \cdot (KL(p,q)+KL(q,p)) \cr 
&+ n \sum_{0\le i,j \le \frac{m}{2}-1}  {m/2-1 \choose i}{m/2-1 \choose
  j} \frac{p^{i+1}q^{j} + p^{j}q^{i+1}}{2} \log \frac{p^{i+1}q^{j} +
  p^{j}q^{i+1}}{p^{i}q^{j+1} + p^{j+1}q^{i}} \cr
\le& m \cdot (KL(p,q)+KL(q,p)) + n \sum_{1 \le k \le m}  m^k p^{k+1} k
\log \frac{p}{q} \cr
\le& O( \gamma f(n)) + np\log\frac{a}{b} \sum_{1 \le k \le m}  k m^k
p^{k} \le O( \gamma f(n)) + np\log\frac{a}{b} \sum_{k=1}^{\infty}  k
(mp)^{k}  \cr 
\le&O( \gamma f(n)) + np\log\frac{a}{b} \sum_{k=1}^{\infty}  
(2mp)^{k}   \cr
\sigma_{\Psi}[L]^2  \le &m( (p+q)(\log \frac{a}{b})^2 + (2-p-q)(\log\frac{1-q}{1-p})^2) \cr &+ n \sum_{0\le i,j \le \frac{m}{2}-1}  {m/2-1 \choose i}{m/2-1 \choose
  j} \frac{p^{i+1}q^{j} + p^{j}q^{i+1}}{2} \left(\log \frac{p^{i+1}q^{j} +
  p^{j}q^{i+1}}{p^{i}q^{j+1} + p^{j+1}q^{i}} \right)^2\cr
\le& 4m p (\log \frac{a}{b})^2 + n \sum_{1 \le k \le m}  m^k p^{k+1} k^2
(\log \frac{a}{b})^2 \cr 
\le& O(\gamma f(n)) + np (\log \frac{a}{b})^2 \sum_{k=1}^{\infty}
k^2 (mp)^{k} \le O(\gamma f(n)) + np (\log \frac{a}{b})^2 \sum_{k=1}^{\infty}
 (3mp)^{k} .
\end{align}
Therefore, the necessary condition for $\mathbb{E}_{\Psi}[L] +
2\sigma_{\Psi}[L] = \omega(1)$ is that $np \sum_{k=1}^{\infty}
 (3mp)^{k} = \omega(1)$. We conclude this proof from that $np \sum_{k=1}^{\infty}
 (3mp)^{k} = \omega(1)$ if and only if $\gamma f(n)^2 = \omega(1)$.

We now prove point (ii). Note that the probability for a red node to
be isolated is at least $(1-af(n)/n)^{\gamma n}\approx \exp(-a\gamma
f(n))$. If there exists an asymptotically accurate clustering
algorithm, then the fraction of such isolated red nodes should
vanishes and hence $\gamma f(n)\to \infty$.

\subsection{Proof of Theorem~\ref{th:4}}

The proof proceeds in two steps. Step 1. We first establish that if $\frac{\sigma^{(g)}_K}{\sqrt{m \hat{p}^{(g)}}} \cdot 1_{\{ m\hat{p}^{(g)} \ge 50\}} = \omega(1)$, then the spectral method applied to the matrix $A^{(g)}$ is asymptotically accurate. We also show that if $\frac{\sigma'_K}{\sqrt{m \hat{p}'}} \cdot 1_{\{ m \hat{p}' \ge 50\}} = \omega(1)$, then the spectral method applied to the matrix of indirect edges $A'$ is asymptotically accurate. Step 2. We show that if $\gamma f(n)=\omega(1)$, then $\frac{\sigma^{(g)}_K}{\sqrt{m \hat{p}^{(g)}}} \cdot 1_{\{ m\hat{p}^{(g)} \ge 50\}} = \omega(1)$ with high probability (w.h.p.), and if $\gamma f(n)=O(1)$ and $\sqrt{\gamma}f(n)=\omega(1)$, then $\frac{\sigma'_K}{\sqrt{m \hat{p}'}} \cdot 1_{\{ m \hat{p}' \ge 50\}} = \omega(1)$ w.h.p.. 

\medskip
\noindent
{\bf Preliminaries.} We first state three lemmas to analyze the performance of Approx, PowerMethod, and Detection algorithms. Their proofs are postponed to the end of this section. In what follows, let $V=\{1,\ldots,n\}$ and let $A\in \mathbb{R}^{n\times n}$. For any matrix $Z\in \mathbb{R}^{n\times n}$, $\lambda_K(Z)$ denotes the $K$-th largest singular value of $Z$. 
 
\begin{lemma}
With probability $1-O(1/n)$, the output $(Q,\sigma_K)$ of the PowerMethod algorithm with input $(A,V,K)$ (Algorithm~\ref{alg:pm}) satisfies that $\sigma_K = \Theta (\lambda_K (A))$.\label{lem:powersing}
\end{lemma}

\begin{lemma} Let $A, M\in \mathbb{R}^{n\times n}$ and let $M=U
  \Lambda U^T$ be the SVD of $M$ where $\Lambda \in\mathbb{R}^{K\times K}$.  Assume that $\|A-M \|= o \left( \lambda_K ( M ) \right)$, the output $(Q,\sigma_K)$ of the PowerMethod algorithm (Algorithm~\ref{alg:pm}) with input $(A,V,K)$ satisfies:
$$
\| U_{\bot}^T Q \| =O\left( \frac{\|A-M \| }{\lambda_K (M)}
\right) = o(1), 
$$
where $U_{\bot}$ is an orthonormal basis of the space perpendicular to the linear span of $U$. \label{lem:poweriter}
\end{lemma}

\begin{lemma} Assume that the set $V$ is partitioned into $K$ subsets $(V_k)_{1\le k\le K}$. Further assume that for any $k$, ${| V_k|\over n}>0$ does not depend on $n$. Let $W$ be the $V \times K$ matrix with for all $(v,k)$, $W_{vk}=1/\sqrt{|V_k|}$ if $v\in V_k$ and $0$ otherwise. Let $W_{\bot}$ be an orthonormal basis of the space perpendicular to the
linear span of $W$. The output $(S_k)_{1\le k \le K}$ of the Detection algorithm with input $(Q,V,K)$ satisfies: if $\| W_{\bot}^T Q \| = o(1)$, then there exists a permutation $\zeta$ of $\{1,\ldots,K\}$ such that
 $$\frac{\left| \bigcup_{k=1}^K S_k
\setminus V_{\zeta(k)} \right|}{n} = O \left( \|W_{\bot}^T Q \|^2\right).$$\label{lem:detection}
\end{lemma}

\medskip
\noindent
{\bf Step 1.} We use the notations introduced in the pseudo-codes of the various algorithms. Let $M^{(g)} = \mathbb{E}[A^{(g)}] $ and $M' = \mathbb{E}[A']$. Let $A^{(g)}_{\Gamma} = (A^{(g)}_{vw})_{v,w\in \Gamma^{(g)}}$ and $M^{(g)}_{\Gamma} = (M^{(g)}_{vw})_{v,w\in
  \Gamma^{(g)}}$. Analogously, we define $A'_{\Gamma}= (A'_{vw})_{v,w\in \Gamma'}$ and $M'_{\Gamma}= (M'_{vw})_{v,w\in \Gamma'}$. 
  
We prove that if $\frac{\sigma'_K}{\sqrt{m \hat{p}'}} \cdot 1_{\{ m \hat{p}' \ge 50\}} = \omega(1)$, then the spectral method applied to the matrix of indirect edges $A'$ is asymptotically accurate. We omit the proof of the asymptotic accuracy of the spectral method applied to $A^{(g)}$ under the condition $\frac{\sigma^{(g)}_K}{\sqrt{m \hat{p}^{(g)}}} \cdot 1_{\{ m\hat{p}^{(g)} \ge 50\}} = \omega(1)$ (since it can be conducted in the same way).

Recall that $\sigma_K'$ denotes the $K$-th largest singular value of the trimmed matrix $A_{\Gamma}'$. Observe that by assumption, for $n$ large enough, $m \hat{p}' \ge 50$. Hence applying the law of large numbers, we can conclude that the largest singular value $\xi_1$ of $A'$ scales at most as $m\hat{p}'$ w.h.p.. Since $\sigma_K'\le \sigma_1'\le \xi_1$ (where $\sigma_1'$ is the largest singular value of $A_\Gamma'$) and $\frac{\sigma'_K}{\sqrt{m \hat{p}'}}= \omega(1)$, we deduce that $m\hat{p}'=\omega(1)$ w.h.p.. Hence the trimming step in the Approx algorithm applied to $(A',\hat{p}',V^{(g)},K)$ does remove a negligible proportion of green nodes, i.e., w.h.p. $| V^{(g)}\setminus \Gamma |=o(| V^{(g)}|)$ or equivalently $| \Gamma' |=m(1+o(1))$.

Observe that w.h.p., $\hat{p}' = \frac{\sum_{u,v} M'_{uv}}{m^2}(1+o(1)) = \Theta( \max_{uv} \{ M_{uv}' \})(1+o(1))$ by the law of large numbers and $\sum_{w \in \Gamma'} A'_{vw} =O(m\hat{p}')$ for all $v\in \Gamma'$. From random matrix theory \cite{feige2005spectral}, with probability $1-O(1/m)$, $\|A'_{\Gamma} - M'_{\Gamma} \| =O(\sqrt{m \hat{p}'})$. Next we apply Lemma \ref{lem:powersing} to $(A_{\Gamma}',\Gamma',K)$ and deduce that $\sigma'_K =\Theta(\lambda_K(A'_{\Gamma}))$ w.h.p.. From $\lambda_K(M'_{\Gamma}) \ge \lambda_K(A'_{\Gamma}) -\|A'_{\Gamma} - M'_{\Gamma} \|$, and $\frac{\sigma'_K}{\sqrt{m \hat{p}'}}  = \omega(1)$, we deduce that w.h.p.,
$$
\frac{\lambda_K(M'_\Gamma)}{\|A'_{\Gamma} -M'_{\Gamma} \|}= \omega(1).
$$
If $M_{\Gamma}'=U\Lambda U^T$, we deduce from Lemma \ref{lem:poweriter} applied to $A_\Gamma'$ and $M_\Gamma'$ that w.h.p., $\| U_{\bot}^T Q \| =o(1)$. We can now apply Lemma \ref{lem:detection} replacing $V$ by $\Gamma'$ and $V_k$ by $\Gamma'\cap V_k^{(g)}$. Observe that the linear span of $U$ coincides with that of $W$ (refer to Lemma \ref{lem:detection} for the definition of $W$). Hence, w.h.p., the nodes $\Gamma'$ are accurately classified, and so are the nodes in $V^{(g)}$.

\medskip
\noindent
{\bf Step 2.}   
We distinguish two cases: 1. $\gamma f(n)=\omega(1)$; 2. $\gamma f(n) = O(1)$ and $\sqrt{\gamma}f(n)=\omega(1)$.

\medskip
\noindent
{\bf Case 1.} Assume that $\gamma f(n)=\omega(1)$. By the law of large numbers, $m\hat{p}^{(g)} = \Theta(\gamma f(n))$ w.h.p..
Since $\lambda_K (M_{\Gamma}^{(g)}) = \Omega(\gamma f(n))$,  $\| A^{(g)}_{\Gamma} - M^{(g)}_{\Gamma}\| =\Theta(\sqrt{m\hat{p}^{(g)}}) = \Theta(\sqrt{\gamma f(n)})$ and $\lambda_{K}(A_{\Gamma}^{(g)}) \ge \lambda_{K}(M_{\Gamma}^{(g)}) -\|A_{\Gamma}^{(g)}-M_{\Gamma}^{(g)} \|$, we get $\frac{\lambda_K (A_{\Gamma}^{(g)})}{\sqrt{ m \hat{p}^{(g)}}} = \omega(1)$ w.h.p.. Since $\sigma^{(g)}_{K} = \Theta (\lambda_K  (A_{\Gamma}^{(g)}))$ from Lemma~\ref{lem:powersing}, w.h.p.
$$
\frac{\sigma^{(g)}_{K}}{\sqrt{m\hat{p}^{(g)}}}\cdot 1_{\{ m \hat{p}^{(g)} \ge 50\}} = \omega(1). 
$$

\medskip
\noindent
{\bf Case 2.} Assume that $\gamma f(n) = O(1)$ and $\sqrt{\gamma}f(n)=\omega(1)$. We first compute $M_{ij}'$ for any $i,j\in V^{(g)}$. For notational simplicity, $\alpha_k=\frac{|V^{(g)}_{k}|}{n}$ and $\beta_k=\frac{|V^{(r)}_{k}|}{n}$. \\
(i) Let $i,j$ be two green nodes belonging to the same community, i.e., $i,j\in V_k^{(g)}$. Let $v\in V^{(r)}_k$. We have:
$$
\mathbb{P}\left[ A_{vi}=1=A_{vj},\sum_{w\in V^{(g)}}A_{vw}=2\right]=p^2(1-p)^{\alpha_kn-2}\prod_{l\neq k}(1-q)^{\alpha_ln}.
$$ 
This probability is equivalent to $p^2\exp(-\alpha_kpn-\sum_{l\neq k}\alpha_lqn)$ when $n\to\infty$. Similarly, when $v\in V_{k'}^{(g)}$ for some $k'\neq k$, the probability $\mathbb{P}[ A_{vi}=1=A_{vj},\sum_{w\in V^{(g)}}A_{vw}=2]$ is equivalent to $q^2\exp(-\alpha_{k'}pn-\sum_{l\neq k'}\alpha_lqn)$ when $n\to\infty$. We deduce that:
\begin{equation}
M_{ij}' \sim p^2 n \beta_k \eta_k  + q^2 n \sum_{k' \neq k} \beta_{k'}\eta_{k'}, \quad\hbox{as }n\to\infty,
\label{eq:mkk}
\end{equation}
where $\eta_k=\exp(-\alpha_kpn-\sum_{l\neq k}\alpha_lqn)$. \\
(ii) Let $i,j$ be two green nodes belonging to different communities, i.e., $i\in V_k^{(g)}$ and $j\in V_{\ell}^{(g)}$, for $k\neq \ell$. Using the same analysis as above, we have:
\begin{equation}
M'_{ij} \sim pq (\beta_k \eta_k  + \beta_\ell \eta_\ell )n  + q^2 n \sum_{k'
  \notin \{k,\ell\}} \beta_{k'} \eta_{k'}\quad\hbox{as }n\to\infty. \label{eq:mij}
\end{equation}
  
From (\ref{eq:mkk})-(\ref{eq:mij}) and the law of large numbers, we get w.h.p., $m\hat{p}'=\Theta(\gamma f(n)^2)$ (this comes from the facts that $\gamma f(n)=O(1)$ and $\alpha_kpn=\Theta(\gamma pn)=\Theta(\gamma f(n))$). As a consequence, $m\hat{p}'=\omega(1)$ w.h.p.. Thus, in the trimming process in the Approx algorithm applied to $(A',\hat{p}',V^{(g)},K)$, we must have $| V^{(g)}\setminus \Gamma'|=o(|V^{(g)}|)$ w.h.p.. 

We also deduce from the above analysis that we can represent $M_{\Gamma}'$ as follows:
$$ M'_\Gamma = M^{(g)}_{\Gamma' , K}\Lambda' (M^{(g)}_{\Gamma' , K})^T ,$$
where $M^{(g)}_{\Gamma' , K}$ is a $\Gamma' \times K$ matrix where
the $k$-th column of $M^{(g)}_{\Gamma' , K}$ is the column vector of
$M^{(g)}_{\Gamma}$ corresponding to $ v \in V^{(g)}_k$, and $\Lambda'$ is a $K\times K$ diagonal
matrix where $k$-th element is $\beta_k
\eta_k n$. Since $\frac{\|M^{(g)}_{\Gamma' , K} {\boldsymbol x}
  \|}{\|{\boldsymbol x}\|} = \Omega(\sqrt{m}p)$ for any ${\boldsymbol x}\in \mathbb{R}^{K\times 1}$, $\lambda_K (M'_{\Gamma}) = \Omega (mnp^2 \min\limits_{1\le k \le
  K} \eta_k ) = \Omega(\gamma f(n)^2)$. By the law of large numbers, w.h.p., $m\hat{p}' = \Theta( \gamma f(n)^2)$. 
Then, as in the analysis of Case 1, we conclude that w.h.p.
$$
\frac{\sigma'_{K}}{\sqrt{m\hat{p}'}}\cdot 1_{\{ m \hat{p}' \ge 50\}} = \omega(1). 
$$
 
 \subsection{Proof of Lemma~\ref{lem:powersing}}
To conclude $\sigma_K = \Theta(\lambda_K(A))$, we show that $\sigma_K
= O(\lambda_K(A))$ (Step 1) and $\sigma_K = \Omega(\lambda_K(A))$
(Step 2).

\medskip
\noindent
{\bf Step 1.} When $\|A\| = \Theta (\lambda_K(A))$, this is trivial since
singular values of $R_{\tau^{\star}}$ have to be less than
$\|A\|$. Let  $\|A\| = \omega (\lambda_K(A))$. Then, there exists
$\ell < K$ such that $\lambda_{\ell}(A) = \omega(\lambda_K(A))$ and $\lambda_{\ell+1}(A) =
\Theta(\lambda_K(A))$. We denote by $\tilde{U}_j \tilde{\Lambda} \tilde{U}_j^T$ be the SVD of rank $j$
approximation of $A$.  Let $Q_{K,\tau^{\star}}$ denote the $K$-th column
vector of $Q_{\tau^{\star}}$. Analogously with Step 1 of the
proof of Lemma~\ref{lem:poweriter}, we can show that
$\|\tilde{U}_{j}^T Q_{K,\tau^{\star}}\| = O(\frac{\lambda_K
  (A)}{\lambda_{j}(A)})$ for all $j \le \ell$. Therefore, $\sigma_{k} = O(\lambda_K(A))$.

\medskip
\noindent
{\bf Step 2.} When $\lambda_{n}(A) = \Theta (\lambda_K(A))$, this is trivial since
singular values of $R_{\tau^{\star}}$ have to be larger than
$\lambda_{n}(A)$. Let $\lambda_{n}(A) = o (\lambda_K(A))$. Then, there
exists $\ell \ge K$ such that
$\lambda_{\ell+1}(A) = o(\lambda_K(A))$ and $\lambda_{\ell}(A) =
\Theta(\lambda_K(A))$. Analogously with Step 1 of the
proof of Lemma~\ref{lem:poweriter}, we can show that
$\|(\tilde{U}_{\ell})_{\bot}^T Q_{K,\tau^{\star}}\| = O(\frac{\lambda_{\ell+1}
  (A)}{\lambda_{\ell}(A)}) = o(1)$, where $(\tilde{U}_{\ell})_{\bot}$ is an
orthonormal basis of the perpendicular to the linear span of
$\tilde{U}_{\ell}$. Therefore, $\sigma_{k} = \Omega (\lambda_\ell (A)) =
\Omega (\lambda_K (A))$.

\subsection{Proof of Lemma~\ref{lem:poweriter}}
We denote by $\tilde{A} = \tilde{U} \tilde{\Lambda} \tilde{U}^T$ be the SVD of rank $K$
approximation of $A$. Let $U_\bot$ and $\tilde{U}_\bot$ be orthonormal bases of
the perpendicular spaces to the linear spans of $U$ and $\tilde{U}$, respectively. 
Since
\begin{align*}
\|U_{\bot}^T Q_{\tau^{\star}} \| & = \|U_{\bot}^T (\tilde{U}\tilde{U}^T +
\tilde{U}_{\bot}\tilde{U}_{\bot}^T )Q_{\tau^{\star}} \| \le \|U_{\bot}^T
\tilde{U}\tilde{U}^T Q_{\tau^{\star}} \| + \|U_{\bot}^T
\tilde{U}_{\bot}\tilde{U}_{\bot}^T Q_{\tau^{\star}} \| \cr
& \le \|U_{\bot}^T\tilde{U}\| \|\tilde{U}^T Q_{\tau^{\star}} \| + \|U_{\bot}^T
\tilde{U}_{\bot} \| \| \tilde{U}_{\bot}^T Q_{\tau^{\star}} \| \le \|U_{\bot}^T\tilde{U}\| +  \| \tilde{U}_{\bot}^T Q_{\tau^{\star}} \|,
\end{align*}
to conclude this proof, we will show that $\| \tilde{U}_{\bot}^T
Q_{\tau^{\star}} \| =O\left( \frac{\|A-M \| }{\lambda_K (M)} \right)$ and $\|U_{\bot}^T\tilde{U}\| = O\left( \frac{\|A-M \| }{\lambda_K (M)}
\right)$.

\medskip
\noindent
{\bf Step 1. $\| \tilde{U}_{\bot}^T
Q_{\tau^{\star}} \| =O\left( \frac{\|A-M \| }{\lambda_K (M)}
\right)$:} Let $\boldsymbol{x}_1$ be the right singular vector of $\tilde{U}_{\bot}^T Q_{\tau + 1}$
corresponding to the largest singular value and
$\tilde{\boldsymbol{x}}_1$ be a $K\times 1$ vector such that $\boldsymbol{x}_1 = R_{\tau+1}\tilde{\boldsymbol{x}}_1$. Then,
\begin{align}
\|\tilde{U}_{\bot}^T Q_{\tau + 1} \|_2^2 &= \frac{\|\tilde{U}_{\bot}^T Q_{\tau + 1}
  \boldsymbol{x}_1 \|_2^2}{\|\boldsymbol{x}_1\|^2_2} =
\frac{\|\tilde{U}_{\bot}^T Q_{\tau + 1} R_{\tau+1}
  \tilde{\boldsymbol{x}}_1 \|_2^2}{\|R_{\tau+1}\tilde{\boldsymbol{x}}_1\|^2_2} \cr&=\frac{\|\tilde{U}_{\bot}^T Q_{\tau + 1} R_{\tau+1}
  \tilde{\boldsymbol{x}}_1
  \|_2^2}{\|\tilde{U}^T Q_{\tau+1} R_{\tau+1}\tilde{\boldsymbol{x}}_1\|^2_2 + \|\tilde{U}_{\bot}^T Q_{\tau+1} R_{\tau+1}\tilde{\boldsymbol{x}}_1\|^2_2} \cr&=\frac{\|\tilde{U}_{\bot}^T AQ_{\tau }
  \tilde{\boldsymbol{x}}_1
  \|_2^2}{\|\tilde{U}^T AQ_{\tau }\tilde{\boldsymbol{x}}_1\|^2_2 +
  \|\tilde{U}_{\bot}^T AQ_{\tau }\tilde{\boldsymbol{x}}_1\|^2_2} \cr
&\le  \frac{\|A -M \|_2^2}{ (\lambda_K (M) - \|A -M \|_2 )^2 (1- \|\tilde{U}_{\bot}^T
  Q_{\tau} \|_2^2) +   \|A -M \|_2^2} , \label{eq:uqbnd}
\end{align}
where the last inequality stems from that
\begin{align*}
\|\tilde{U}_{\bot}^T AQ_{\tau }  \tilde{\boldsymbol{x}}_1  \|_2  
&\le \| \tilde{U}_{\bot}^T A\| \|Q_{\tau } \| \| \tilde{\boldsymbol{x}}_1
\|_2 \le \| \tilde{U}_{\bot}^TA \|  \| \tilde{\boldsymbol{x}}_1
\|_2  = \lambda_{K+1}(A) \| \tilde{\boldsymbol{x}}_1\|_2 \le\| (A - M)\| \|\tilde{\boldsymbol{x}}_1\|_2 \cr
\|\tilde{U}^T AQ_{\tau }\tilde{\boldsymbol{x}}_1\|_2 & = \|(\tilde{U}^T
A \tilde{U} \tilde{U}^T Q_{\tau } + \tilde{U}^T A \tilde{U}_\bot
\tilde{U}_{\bot}^T Q_{\tau })\tilde{\boldsymbol{x}}_1\|_2 =  \|\tilde{U}^TA \tilde{U} \tilde{U}^T Q_{\tau }
\tilde{\boldsymbol{x}}_1 \|_2 \cr
& \ge \lambda_K(A) \|\tilde{U}^TQ_{\tau}\tilde{\boldsymbol{x}}_1\|_2 \ge (\lambda_K(M) - \|A-M \| ) \|\tilde{U}^TQ_{\tau}\tilde{\boldsymbol{x}}_1\|\cr 
& \ge (\lambda_K(M) - \|A-M \| ) \sqrt{1- \|\tilde{U}_{\bot}^T
  Q_{\tau} \|_2^2} \| \tilde{\boldsymbol{x}}_1 \|.
\end{align*}

Let $\zeta = \frac{\|A -M \|_2^2}{(\lambda_K (M) -
  \|A -M \|_2 )^2}$.  Since $\frac{\|A -M
  \|_2}{\lambda_K (M)} = o(1)$, $\zeta= O(\frac{\|A -M \|_2^2}{\lambda_K (M) ^2}) = o(1)$. Then, from \eqref{eq:uqbnd},
$$1-\|\tilde{U}_{\bot}^T Q_{\tau + 1} \|_2^2 \ge 1-\frac{\zeta}{1-\|\tilde{U}_{\bot}^T
  Q_{\tau} \|_2^2 +\zeta} = \frac{1-\|\tilde{U}_{\bot}^T
  Q_{\tau} \|_2^2 }{1-\|\tilde{U}_{\bot}^T
  Q_{\tau} \|_2^2 +\zeta}.$$
When $1-\|\tilde{U}_{\bot}^T  Q_{\tau} \|_2^2 \le \zeta$, $1-\|\tilde{U}_{\bot}^T Q_{\tau + 1} \|_2^2 \ge  \frac{1-\|\tilde{U}_{\bot}^T
  Q_{\tau} \|_2^2 }{2\zeta}.$ From this, one can easily check that when $\tau \ge \frac{\log
  (\zeta/(1-\|\tilde{U}_{\bot}^T Q_{0}\|_2^2))}{\log (1/2 \zeta)}$,
$1-\|\tilde{U}_{\bot}^T  Q_{\tau} \|_2^2 \ge \zeta$,
$1-\|\tilde{U}_{\bot}^T  Q_{\tau+1} \|_2^2 \ge 1/2$, and
$1-\|\tilde{U}_{\bot}^T  Q_{\tau+2} \|_2^2 \ge 1- 2\zeta$.
Therefore, when $\frac{\log |V|}{2} \ge \frac{\log
  (\zeta/(1-\|\tilde{U}_{\bot}^T Q_{0}\|_2^2))}{\log (1/2 \zeta)}$, $\|\tilde{U}_{\bot}^T Q_{\tau^{\star}} \|_2=O\left( \frac{\|A-M \| }{\lambda_K (M)}
\right)$. Since $\zeta = o(1)$, to complete this proof, it is sufficient to show that $1-\|\tilde{U}_{\bot}^T
Q_{0}\|_2^2 \ge 1/\mbox{Poly}(n)$ with probability $1-O(1/n)$, where $\mbox{Poly}(n)$
is a polynomial function of $n$ with finite
order. By Theorem 1.2 of \cite{rudelson2008} (Please refer
to the proof of Lemma 10 of \cite{caramanis2013}) we can conclude this part with $\mbox{Poly}(n)=1/n^4$.

\medskip
\noindent
{\bf Step 2. $\|U_{\bot}^T\tilde{U}\| = O\left( \frac{\|A-M \| }{\lambda_K (M)}
\right)$:} We can get an upper bound and a lower bound for $\|  A U_{\bot} \|$ as follows:
\begin{align*}
\|  A U_{\bot} \|& = \|  (M + A - M) U_{\bot} \| =
\|  (A - M) U_{\bot} \| \le \|A - M \| \cr
\|  A U_{\bot} \|& = \|  (\tilde{A} + A - \tilde{A}) U_{\bot} \| \ge
\|  \tilde{A}U_{\bot} \|- \|  (A - \tilde{A})
U_{\bot} \|  \ge
\|  \tilde{U} \tilde{\Lambda} \tilde{U}^T U_{\bot} \|- \| A -
\tilde{A} \| \cr & \ge \lambda_{K}(A) \| \tilde{U}^T U_{\bot} \|- \| A
- \tilde{A} \| \ge ( \lambda_{K}(M) - \|A-M\|) \| \tilde{U}^T U_{\bot} \|- \| A - M \| .
\end{align*}
When we combine above bounds, 
$ \| \tilde{U}^T U_{\bot} \| \le \frac{2 \| A - M \|}{ ( \lambda_{K}(M) - \|A-M\|)} =O\left( \frac{\|A-M \| }{\lambda_K (M)}
\right)$.

\subsection{Proof of Lemma~\ref{lem:detection}}\label{sec:proof-lemma-refl-1}
From the definitions of $W $ and $W_{\bot}$,
$Q = W W^T Q + W_{\bot}W_{\bot}^TQ$.
Since rows of $W$ corresponding to the nodes from the same cluster
are the same, the rows of $W W^T Q$ are also the same for the node from
the same clusters. Let $W W^T Q(k)$ be
the rows of $W W^T Q$ corresponding to $v\in V_k$. Let
${\boldsymbol v}_{(k \ell)} \in
\mathbb{R}^{K\times 1}$ such that $k$-th row and $\ell$-th row are $1/\sqrt{|V_k|}$ and $-1/\sqrt{|V_\ell|}$,
respectively and other elements  are zero. Then, $\| W W^T Q(k) - W W^T Q(\ell)
\|^2 = \| W^T Q {\boldsymbol v}_{(k \ell)}\|^2
$.  Since $\|W^T Q
\boldsymbol{x} \| \ge \sqrt{1 - \| W_{\bot}^TQ \|^2}\|\boldsymbol{x} \|$, 
$$\| W W^T Q(k) - W W^T Q(\ell)
\|^2 = \Omega( \frac{1 - \| W_{\bot}^TQ \|^2}{n} )= \Omega(\frac{1}{n}) \quad\mbox{for all}~
k\neq \ell. $$  
Therefore, with some positive $C>0$,
\begin{align*}
C &\frac{ |\bigcup_{k,\ell: k \neq \ell}
  S_k \bigcap V_\ell |}{n}  \le \sum_{k,\ell: k \neq \ell}
\sum_{v\in  S_k \bigcap V_\ell} \|W W^T Q(k) - W W^T Q(\ell ) \|^2 \cr
& \le 2 \sum_{k,\ell : k \neq \ell}
\sum_{v\in  S_k \bigcap V_\ell} \|W W^T Q(k) - \xi_{i^{\star},k} \|^2
+\|\xi_{i^{\star},k} - W W^T Q(\ell ) \|^2 \cr
&\le 4 \sum_{k,\ell : k \neq \ell}
\sum_{v\in  S_k \bigcap V_\ell} \| W W^T Q(\ell ) - \xi_{i^{\star},k}
\|^2 \cr
&\le 8 \sum_{k,\ell : k \neq \ell}
\sum_{v\in  S_k \bigcap V_\ell} \| W W^T Q(\ell ) - Q_v
\|^2 + \|  Q_v - \xi_{i^{\star},k} \|^2 \cr
&\le 8  \|W_{\bot}W_{\bot}^TQ\|_F^2 + 8r_{i^{\star}} \le 8 K
\|W_{\bot}W_{\bot}^TQ\|^2 + 8r_{i^{\star}} \le 8 K
\|W_{\bot}^TQ\|^2 + 8r_{i^{\star}}
\end{align*}

To conclude this proof, we need to show that $r_{i^{\star}} =
O(\|W_{\bot}^TQ\|^2)$. Let $i^t$ be an integer between $1$ and $\log
n$ such that  $\frac{\|W_{\bot}W_{\bot}^TQ\|_F^2 }{n
  \delta^2} \le \frac{i^t}{n \log n} \le \frac{\delta^2}{n}$ with
positive constant
$\delta$ close to 0. There exists such $i^t$ for any $\delta$, since   $\|W_{\bot}W_{\bot}^TQ\|^2 = o(1)$ and the rank of
$W_{\bot}W_{\bot}^TQ$ is $K$. Then, 
\begin{align*}
\Big|\bigcup_{1\le k \le K} \{ v \in V_k : \|Q_v - W W^T Q(k ) \|^2 \le
\frac{i^t }{4 n \log n} \} \Big| &\ge n -
\|W_{\bot}W_{\bot}^TQ\|_F^2\frac{4 n\log n}{i^t } \cr
&\ge n(1-4\delta^2).\end{align*}
From this, since $\|Q_v - Q_w \|^2 \le 2\|Q_v - W W^T Q(k )\|^2 + 2\|Q_w -
W W^T Q(k )\|^2 $, when $v$ satisfying that $\|Q_v - W W^T Q(k ) \|^2 \le
\frac{i^t}{4 n\log n},$ $$|X_{i^t,v}| \ge |V_k|- 4\delta^2 n.$$ On
the other hand, since $\|Q_v - Q_w \|^2 \ge \frac{1}{2}\|Q_v - W W^T Q(k )\|^2 -\|Q_w -
W W^T Q(k )\|^2 $, when $v$ satisfying that $\|Q_v - W W^T Q(k ) \|^2 \ge
\frac{i^t}{ n\log n},$ $$|X_{i,v}| \le  4\delta^2 n.$$ With small
enough constant $\delta$, therefore, when $v$ and $w$ satisfy that
$\|Q_v - W W^T Q(k ) \|^2 \le
\frac{i^t}{4 n \log n}$ and $\|Q_w - W W^T Q(k ) \|^2 \ge
\frac{i^t}{ n \log n},$ $|X_{i,v}| > |X_{i,w}|$, which indicates
that the origin of $T_{i^t, k}$ is at least $\|Q_{v_k} - W W^T Q(k ) \|^2 \le
\frac{i^t}{ n \log n}$ and $|T_{i^t,
, k}| \ge |V_k| - 4\delta^2 n $. Since $\|\cdot \|$ is a
convex function, by Jensen's inequality, for all $k$, 
$$\|W W^T Q(k ) - \xi_{i^t,k}\|^2 \le \frac{\sum_{v \in T_{i^t , k}}
  \|W W^T Q(k ) - Q_v\|^2 }{|T_{i^t , k}|} \le
\frac{\|W_{\bot}W_{\bot}^TQ\|_F^2 }{|V_k| - 4\delta^2 n} =  O(\frac{\|W_{\bot}^TQ\|^2 }{n}).$$
Therefore,
\begin{align*}
r_{i^t} & =  \sum_{k=1}^K \sum_{v \in T_{i^t,k}} \|Q_v - \xi_{i^t,k} \|^2  \le \sum_{k=1}^K \sum_{v \in V_k} \|Q_v - \xi_{i^t,k} \|^2 \cr
 &\le 2\sum_{k=1}^K \sum_{v \in V_k } \|Q_v - W W^T Q(k ) \|^2 +\|W W^T
 Q(k ) - \xi_{i^t,k} \|^2 \cr
 &\le 2 \|W_{\bot}W_{\bot}^TQ\|_F^2 +2\sum_{k=1}^K \sum_{v \in V_k }
 \|W W^T Q(k ) - \xi_{i^t,k} \|^2 \cr
& = O ( \|W_{\bot}^TQ\|^2)+2\sum_{k=1}^K \sum_{v \in V_k }
 \|W W^T Q(k ) - \xi_{i^t,k} \|^2 = O ( \|W_{\bot}^TQ\|^2).
\end{align*}
Since $r_{i^{\star}} \le r_{i^t}$, $r_{i^{\star}} = O ( \|W_{\bot}^TQ\|^2).$

\subsection{Proof of Theorem~\ref{thm:clust-algor-red}}
Let $\mu ( v, S^{(g)}_k ) =
\mathbb{E} [ \sum_{w \in S^{(g)}_k} A_{vw}]$ and $Var ( v, S^{(g)}_k ) =
\mathbb{E} [ (\mu ( v, S^{(g)}_k ) - \sum_{w \in S^{(g)}_k} A_{vw})^2]$. Since $|\bigcup_{k=1}^K (S^{(g)}_k
\setminus V^{(g)}_k ) | =o(|V^{(g)}|)$ from Theorem~\ref{th:4}, $\frac{\mu ( v,
  S^{(g)}_k )}{|S^{(g)}_k|} = p(1+ o(1))$ and $\frac{Var ( v,
  S^{(g)}_k )}{|S^{(g)}_k|} = p(1+ o(1))$ when $v\in V_k$, and $\frac{\mu ( v,
  S^{(g)}_k )}{|S^{(g)}_k|} = q(1+ o(1))$ and $\frac{Var ( v,
  S^{(g)}_k )}{|S^{(g)}_k|} = q(1+ o(1))$ when $v\notin V_k$. By
Chebyshev's inequality, when $v \in V_k$, $v \in S_k$ with
high probability since
$\frac{\mu(v,S^{(g)}_k)-\mu(v,S^{(g)}_{k'})}{\sqrt{Var(v,S^{(g)}_k)}} =
\omega(1)$ for all $k' \neq k$ when $\gamma f(n) = \omega(1)$.

\subsection{Proof of Theorem~\ref{thm:streaming-with-green}}
In this proof, we use Chernoff bound as the form of Lemma 8.1 in \cite{coja2010}.

From Theorem~\ref{th:4}, 
Algorithm~1 classifies the arrival nodes at each time
block with diminishing fraction of misclassified nodes. Between
$S^{(\tau)}_i$ and $S^{(\tau+1)}_j$, the number of connections is
$\Theta( B^2 \frac{\bar{f}(n)}{n}) = \Theta( \frac{h^2(n)n}{\min\{f(n),n^{1/3} \} \log^2 n}) = \omega(1)$ from
the condition of this theorem. Let $\mu(k,i) = \frac{ \sum_{v \in \hat{V}_i}\sum_{w \in
  S^{(\tau)}_{k}} A_{wv}}{ \sum_{v \in \hat{V}_i}\sum_{w \in
  S^{(\tau)}_{k}} 1}$. By the Chernoff bound,
with high probability (since $\sum_{v \in \hat{V}_i}\sum_{w \in
  S^{(\tau)}_{k}} A_{wv} =\omega(1)$),  $\mu(k,i) =p(1-o(1))$ when $\frac{|S^{(\tau)}_k
  \bigcap V_i|}{|S^{(\tau)}_k|} = 1-o(1)$ and $\mu(k,i) =q(1+o(1))$ when $\frac{|S^{(\tau)}_k
  \bigcap V_i|}{|S^{(\tau)}_k|} = o(1)$. Therefore, with high
probability, $S^{(\tau)}_k$ is merged with $\hat{V}_{s(k)}$ such that $\frac{|S^{(\tau)}_k \bigcap V_{s(k)}|}{|S^{(\tau)}_k|} =
1-o(1)$. Thus, $\frac{|\hat{V}_k \bigcap V_{k}|}{|\hat{V}_k|} =
1-o(1)$ for all $k$ with high probability. 

Since $\frac{|\hat{V}_k \bigcap V_{k}|}{|\hat{V}_k|} =
1-o(1)$ for all $k$, one can easily show using the Chernoff bound that $\frac{N_{v,k}}{|\hat{V}_k|} \ge p(1-\frac{p-q}{4})$ when $v
\in V_k$ and $\frac{N_{v,k'}}{|\hat{V}_{k'}|} \le q(1+\frac{p-q}{4})$ when $v
\notin V_{k'}$ with probability $1-O(\exp (-cT\frac{f(n)}{n}) )$ with a constant
$c>0$. Thus, the 
probability for that $\frac{N_{v,k}}{|\hat{V}_{k}|} \le
\frac{N_{v,k'}}{|\hat{V}_{k'}|}$ for $v\in V_k$ and $k \neq k'$ is $O(\exp(-cT\frac{f(n)}{n}))$.

\end{document}